\documentclass[prb,aps,twocolumn,superscriptaddress]{revtex4-2}
\usepackage{graphicx,color}
\usepackage{amsthm}
\usepackage{amsfonts}
\usepackage{algorithmic}
\usepackage{enumerate}
\usepackage{latexsym}
\usepackage{amsmath}
\usepackage{amssymb}
\usepackage[colorlinks=true,citecolor=blue,linkcolor=blue]{hyperref}

\usepackage[T1]{fontenc}

\newcommand{\tcr}{\textcolor{black}}

\begin{document}
\title{Scaling of energy and power in a large quantum battery-charger model}

\author{Lei Gao}
\affiliation{Beijing Computational Science Research Center, Beijing 100193, China}

\author{Chen Cheng}
\affiliation{School of Physical Science and Technology $\&$ Lanzhou Center for Theoretical Physics $\&$ Key Laboratory of Theoretical Physics of Gansu Province, Lanzhou University, Lanzhou, Gansu 730000, China}

\author{Wen-Bin He}
\affiliation{Beijing Computational Science Research Center, Beijing 100193, China}

\author{Rubem Mondaini}
\affiliation{Beijing Computational Science Research Center, Beijing 100193, China}

\author{Xi-Wen Guan}
\affiliation{State Key Laboratory of Magnetic Resonance and Atomic and Molecular Physics, Wuhan Institute of Physics and Mathematics, APM, Chinese Academy of Sciences, Wuhan 430071, China}
\affiliation{Department of Theoretical Physics, RSPE, Australian National University, Canberra, ACT 0200, Australia.}

\author{Hai-Qing Lin}
\email{haiqing0@csrc.ac.cn}
\affiliation{Beijing Computational Science Research Center, Beijing 100193, China}

\begin{abstract}
We investigate a multi-qubit quantum battery-charger model, focusing on its potential emulation on a superconducting qubit chip. Using a large-spin representation, we first obtain the analytical form of the energy $E_B(t)$, power $P_B(t)$ and their maximum values, $E_B^{\rm max}$ and $P_B^{\rm max}$, of the battery part by means of the antiferromagnetic Holstein-Primakoff (AFM-HP) transformation within the low-energy approximation. In this case, our results show that superextensive scaling behavior of $P_B^{\rm max}$ ensues. By further combining these with the ones obtained via exact diagonalization (ED), we classify the dynamics of various physical quantities, including the entanglement between the battery and charger parts for system sizes encompassing over 10,000 qubits. Finally, by checking a diverse set of system configurations, including either a fixed battery size with a growing number of charger qubits, or when both parts simultaneously grow, we classify the system size scalings of $E_B^{\rm max}$ and $P_B^{\rm max}$, relating it with the entanglement entropy in the system. In agreement with the analytical results, robust superextensive behavior of $P_B^{\rm max}$ is also observed in this case. Our work provides an overall guide for expected features in experiments of quantum batteries emulated in superconducting qubit platforms, in particular ones that exhibit long-range couplings. 
\end{abstract}

\maketitle
\section{Introduction \label{sec:introduction}}

Recent breakthroughs in quantum technologies have highlighted the concomitant effort of both theory and experiments in bringing to light phenomena that can surpass capabilities often associated with classical systems. Among those, advances in quantum communications~\cite{ET_2001_NATURE_Zoller,ET_2020_NatPho_Pan,ET_2020_NATURE_Jian-Wei,ET_2021_NATURE_Pan} and quantum computing~\cite{ET_2010_NATURE_Brien,T_2017_PNAS_Matthias,ET_2021_NatPhy_Robin} reveal concrete advantages in contrast to their classical analogues, which leverage on special properties of the quantum realm, as entanglement, coherence and quantum correlations. Combining all three allows experimentalists to perform quantum simulations, via the emulation of elusive physical models using cold atoms~\cite{TE_2021_SCIENCE_Bloch,ET_2019_SCIENCE_Bark,ET_2020_SCIENCE_Jendrzejewski,ET_2020_NATURE_Thompson,ET_2021_NATURE_Lukin},
trapped ions~\cite{E_2011_SCIENCE_Lanyon,E_2012_NATURE_Blatt,E_2019_NatComm_Fabian,E_2020_PRL_Tamura,R_2021_RMP_Monroe}, and superconducting qubit chips~\cite{TE_2021_NatPhy,ET_2021_ScienceBulletin_Yu,ET_2020_SCIENCE_Zalcman,ET_2019_SCIENCE_Pan}.

Coherent preparation and control of a quantum system also endow the ability to tackle the prospects of energy storage using quantum devices. This gave birth to the idea of a quantum battery, a quantum-mechanical system that permits the deposition and extraction of energy, with a claimed performance that overcomes their classical analogues. It was first proposed and demonstrated by Alicki and Fannes~\cite{T_2013_PRE_Fannes}, which by using global entanglement operations, described the maximal amount of extractable work from an isolated quantum system. Doing the reverse process, charging a quantum battery, Ref.~\cite{E_2015_NJP_Binder} reported a similar $N$-fold increase in charging power of the system with global operations, where $N$ is the number of the two-level cells of the battery, originally non-interacting. As a consequence of the difficulties in realizing global entanglement operations in many-particle systems, ideas of quantum batteries with collective, but local, interactions among the cells came to fruition. It has been reported~\cite{E_2017_PRL_Kavan} that the collective behavior of an interacting quantum battery would lead to a quantum advantage in charging power even in the absence of quantum entanglement, obtaining a generic upper bound under a restricted interaction order. Such upper bound has been recently put under even more stringent bounds~\cite{Gyhm2022}. 

The observation of advantages of charging power in an interacting spin-chain model of a many-body quantum battery over its noninteracting correspondent was also investigated~\cite{T_2017_PRA_Felix} -- these originated from the interactions, but however, could be explained on a mean-field level rather than by the correlations between the spins. Enhancements of charging power were also reported in Dicke-like quantum optical models but debate ensued of whether such advantages originated from quantum entanglement~\cite{T_2018_PRL_Marco} or just the coherent cooperative interactions without many-particle entanglement~\cite{T_2018_AXRIV_blaauboer}. 
By comparing quantum batteries with their `classical' analogues defined by standard classical Hamiltonian mechanics, Ref.~\cite{T_2019_PRB_Andolina2} reported that the origin of quantum advantages is model-dependent.
Moreover, focusing on the energy extracted from quantum batteries, Ref.~\cite{T_2019_PRL_Andolina3} claimed that correlations among elements suppressed the extractable energy, which could be mitigated by preparing the charger in a coherent optical state.

More recently, it was reported that much better work extraction capabilities could be also achieved by making use of  low-entangled many-body localized states in comparison to highly-entangled ergodic ones, which paved the way for exploring a disordered quantum battery~\cite{T_2019_PRB_Polini}. Another recent breakthrough advanced rigorous bounds in the charging power, showing that in various previously studied prototypical models for quantum batteries, including integrable spin chains, the Lipkin-Meshkov-Glick model (with infinite-range two-body interactions) or the Dicke model, they all do not show superextensive behavior if correctly characterizing the thermodynamic limit~\cite{T_2020_PRR_Maciej}. Yet, recent exploration of the all-to-all coupled Sachdev-Ye-Kitaev (SYK) quantum battery, with a well-defined thermodynamic limit, avoid this fate and ultimately recovers the quantum advantage in the charging power~\cite{T_2020_PRL_Polini}. 

At the same time, studies that tackle more realistic situations, e.g., that take into account open quantum systems containing interactions with the environment, were also investigated in order to explain how these unavoidable factors affect the performance of quantum batteries~\cite{T_2019_PRB_Farina,T_2019_PRL_Felipe,T_2020_NJP_Ferraro,T_2020_NJP_Sassetti,T_2020_PRA_Jun-Hong,T_2020_PRL_Adolfo,T_2020_PRApplied_Quach,T_2021_PRA_Aditi}. Finally, recent experimental proposals that realize some of these ideas have been put forward in various settings~\cite{E_2021_AXRIV_Yu,Virgili2022}. 

\begin{figure}[h]
    \includegraphics[width=0.45\textwidth]{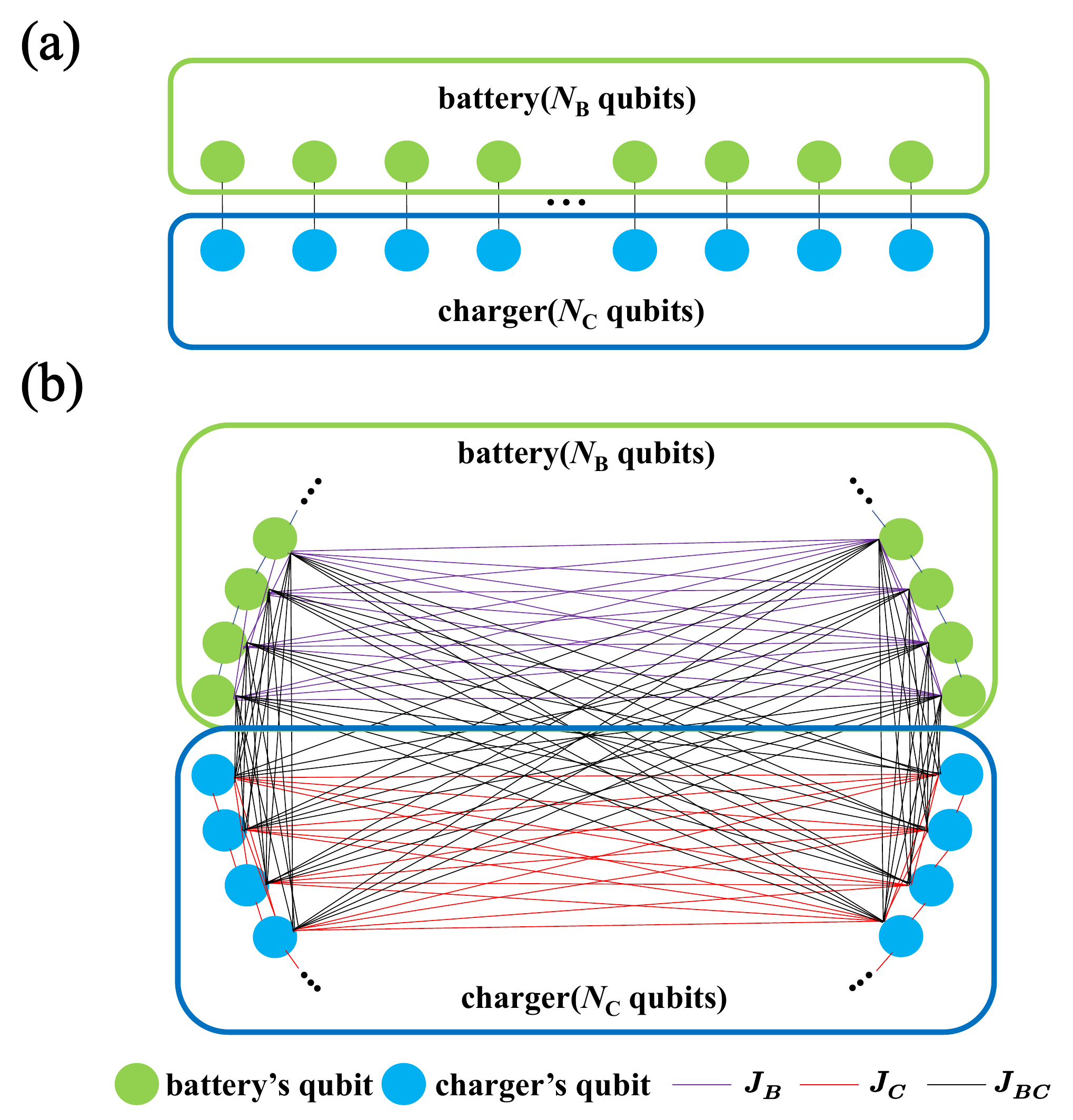}
    \caption{Illustration of the (a) classical (parallel) and (b) quantum (collective) batteries. The $N$ qubits are divided into two parts: the battery
    part (green box) with $N_B$ qubits and charger part (blue box) with $N_C$ qubits. In the parallel case, $N_B=N_C$, and only individual couplings $J_{\rm pair}$
    between the battery and charger qubits exist. In the collective case, couplings $J_B$ among the battery qubits, $J_C$ among the charger qubits and $J_{BC}$ between the battery and charger qubits.}
    \label{FIG1}
    \end{figure}

To clarify the connections between the energy transfer and entanglement in quantum batteries, and provide guidance to upcoming experiments, we consider battery and charger on the same footing in the \textit{absence of external fields}, similar to Ref.~\cite{T_2018_PRB_Andolina} but containing more elements both in the battery and charger part to show collective behaviors\tcr{; a similar idea has also been investigated in Ref.~\cite{T_2021_PRE_Jing}, within the low-energy approximation, and contrasting the advantages in comparison to a charging process promoted by continuous cavity modes.} Furthermore, our aim is to investigate an interacting model that closely resembles recently constructed noisy intermediate-scale quantum (NISQ) devices featuring coherent superconducting qubits~\cite{Xu2020,TE_2021_NatPhy}. These display largely tunable inter-qubit couplings and serve as the basis of our analysis. With the focus on future devices with a large number of elements, we investigate battery-charger models with a much larger number of qubits ($N_B$ and $N_C$) than the currently available NISQ devices (see Fig.~\ref{FIG1}), while pointing out how our results modify for a much smaller number of qubits.

Our presentation is divided as follows: In Sec.~\ref{sec:model}, we introduce the model, further showing analytical results for either a `parallel' or a collective one under the assumption of uniformity of the couplings and onsite potentials. The latter is extracted after applying an anti-ferromagnetic (AFM) Holstein-Primakoff (HP) transformation within the low-energy approximation. Section~\ref{sec:results} displays exact numerical results obtained via exact diagonalization (ED) for various settings of the Hamiltonian parameters, also analyzing the dependence of the charger and total system size. We summarize our results in Sec.~\ref{sec:conclusions}.

\section{Model and Analytical Results \label{sec:model}}

\indent
We investigate a quantum battery-charger model featuring $N$ qubits, as schematically shown in Fig.~\ref{FIG1}(b), being inspired by the actual NISQ device used in Ref.~\cite{TE_2021_NatPhy}. Its Hamiltonian consists of three terms:
\begin{eqnarray}
\hat H = \hat H_B + \hat H_C + \hat H_I\ .
\label{totalHamiltonian}
\end{eqnarray}
The first (second) term describes all the diagonal elements in the Hamiltonian of the $N_B$ ($N_C$) qubits of the battery (charger) part, whereas the third term denotes the (off-diagonal) all-to-all couplings among each pair of the qubits of the system by which energy can be transferred between the two parts, as well as within the battery and charger part separately. Individually, the terms read as,
\begin{equation}
\hat H_{B(C)} = \sum_{\stackrel{m\in N_B}{(\scriptscriptstyle m\in N_C)}\ } V_m \hat \sigma_m^+\hat\sigma_m^- ,
\label{eq:HB_C}
\end{equation}
with interacting part,
\begin{equation}
\hat H_I = \sum_{m\in N, n\in N, m<n} J_{mn}(\hat \sigma_m^+\hat \sigma_n^-+\hat \sigma_m^-\hat \sigma_n^+) .
\label{eq:HI}
\end{equation}
Here, $m$ and $n$ label the qubits, $V_m$ is the onsite potential of qubit $m$, and $J_{mn}$ is the coupling strength between the qubit $m$ and $n$. $\hat \sigma^{\pm}=\frac{1}{2}(\hat\sigma^x \pm {\rm i}\hat \sigma^y)$ are the raising and lowering operators and $\hat \sigma^{x,y,z}$ are the spin-1/2 Pauli matrices.
Focusing on the charging process, we set the initial state as the one in which the battery (charger) is in its lowest-energy (highest-energy) product state. The total initial state of the system $\left|\psi (0)\right>$ is then written as,
\begin{eqnarray}
 \left|\psi (0)\right>=\left|\downarrow \downarrow\dotsb\downarrow \right>_B\bigotimes \left|\uparrow \uparrow\dotsb\uparrow\right>_C\equiv\left|\Downarrow \right>_B\bigotimes \left|\Uparrow\right>_C
\label{eq:initial}.
\end{eqnarray}

Assuming that the system is isolated from the environment, experiencing thus unitary time evolution, the state at time $t$ is $\left|\psi (t) \right>=e^{-{\rm i}\hat Ht}\left|\psi (0) \right>$. The performance of the quantum battery is monitored by the instantaneous energy $E_B(t)$ and average power $P_B(t)$ of the battery part,
\begin{eqnarray}
E_B(t)=\langle\psi(t)|\hat H_B|\psi(t)\rangle,\ 
P_B(t)=\frac{E_B(t)-E_B(0)}{t},  
\label{eq:E_B_P_B}
\end{eqnarray}
as well as their maximum values $E_B^{\rm max}$ and $P_B^{\rm max}$ over time $t$,
\begin{equation}
E_B^{\rm max}=\max\limits_{t}[E_B(t)],\quad
P_B^{\rm max}=\max\limits_{t}[P_B(t)].
\end{equation}
To understand the connections between $E_B(t)$ and $P_B(t)$ with entanglement, we further investigate the von Neumann entropy $S_{\rm vN}$, which reveals how the battery and charger part entangle in the dynamics,
\begin{equation}
S_{\rm vN}=-{\rm Tr}_B\left[\rho_B\log(\rho_B)\right]=-\sum_{i}\lambda_i\log(\lambda_i),
\label{eq:entropy}
\end{equation}
where $\rho_B$ is the reduced density matrix of the battery part, and $\lambda_i$ is the $i$-th eigenvalue of $\rho_B$.

Before systematically studying the properties of the quantum battery, one usually defines a `classical' reference, \emph{i.e.}, a parallel charging battery schematically represented in Fig.~\ref{FIG1}(a), consisting of identical qubit pairs (one qubit in the battery and another in the charger) featuring only intra-pair interactions, as to contrast the differences that the collective interactions will bring. Since there are no interactions between pairs, and each pair experiences its charging processes individually, the energy and power of the parallel battery are thus extensive, that is, proportional to the number of qubits of the whole system. In what follows, we first obtain the analytical expressions for the physical quantities of the parallel battery, where we subsequently generalize them to the collective (non-parallel) case.

\subsection{Parallel battery}

\indent Assuming that the couplings within each pair are homogeneous, the total Hamiltonian of the parallel battery $\hat H^{\parallel}$ can be written in terms of the number of battery-charger pairs $N_B$, i.e., $\hat H^{\parallel}=N_B \hat H^{\rm pair}$, in which $\hat H^{\rm pair}$ is the Hamiltonian of a single pair defined as
\begin{equation}
\hat H^{\rm pair}=J_{\rm pair}(\hat \sigma_b^+\hat \sigma_c^-+\hat\sigma_b^-\hat\sigma_c^+)+V_b\hat \sigma_b^+\hat\sigma_b^-+V_c\hat\sigma_c^+\hat\sigma_c^-,
\label{eq:pair}
\end{equation}
where $J_{\rm pair}$ is the coupling of the two qubits of the pair, $V_{b(c)}$ is the corresponding onsite potential of each qubit of the pair and $N=2N_B=2N_C$; $\hat \sigma_b$ and $\hat \sigma_c$ are the Pauli matrices of the battery and charger qubit of each pair.  

The analytical form of the energy $E_B^{\parallel}(t)=N_B\langle\psi(t)|\hat H^{\rm pair}|\psi(t)\rangle$ thus reads
\begin{equation}
 E^{\parallel}_B(t)=\frac{2N_BV_bJ_{\rm pair}^2}{\Omega^2}\left[1-\cos(\Omega t)\right],
\label{eq:para-energy}
\end{equation}
where $\Omega\equiv\sqrt{4J_{\rm pair}^2+\Delta V^2}$ and $\Delta V\equiv V_b-V_c$. Here, its maximal value being
\begin{equation}
    E_B^{\parallel, {\rm max}}=\frac{4N_BV_bJ_{\rm pair}^2}{\Omega^2},\quad t_E^{\parallel, {\rm max}}=\frac{\pi}{\Omega},
\end{equation}
where $t_E^{\parallel, {\rm max}}$ is the shortest time to reach the $E_B^{\parallel, {\rm max}}$. The corresponding average power for this parallel case $P^{\parallel}_B(t)=E^{\parallel}_B(t) /t$ is written as
\begin{equation}
P^{\parallel}_B(t)=\frac{2N_BV_bJ_{\rm pair}^2}{\Omega^2t}\left[1-\cos(\Omega t)\right], 
\end{equation}
whose maximal value reads
\begin{equation}
    P^{\parallel, {\rm max}}_B\simeq\frac{1.44N_BV_bJ_{\rm pair}^2}{\Omega},\quad t_P^{\parallel, {\rm max}}\simeq\frac{2.33}{\Omega},
\end{equation}
where, similarly, $t_P^{\parallel, {\rm max}}$ is the shortest time to reach the $P^{\parallel, {\rm max}}_B$; one sees that $P_B^{\parallel}(t)$ reaches its maximum faster than $E_B^{\parallel}(t)$ within the same set of parameters. In turn, the corresponding von Neumann entropy of the parallel battery $S^{\parallel}_{\rm vN}$ is
\begin{equation}
S^{\parallel}_{\rm vN}=-N_B\left[A\log(A)+B\log(B)\right],
\label{eq:para-von}
\end{equation}
where $A=\cos^2(\frac{\Omega t}{2})+\frac{\Delta V^2}{\Omega^2}\sin^2(\frac{\Omega t}{2})$ and $B=\frac{4J_{\rm pair}^2}{\Omega^2}\sin^2(\frac{\Omega t}{2})$.
\tcr{These results are further explored in Appendix~\ref{app:parallel} accompanied by an} illustration of the dynamics. As expected, these quantities oscillate in time, but a nonzero $\Delta V$, i.e., introducing an offset between the onsite potentials of the battery and charger qubits of the pair hampers the energy transfer, and the battery part cannot be fully charged under this condition. Additionally, in the absence of an offset, $P_B^{\parallel, {\rm max}}$ is proportional to the coupling of the pair: Improving the magnitude of the coupling $J_{\rm pair}$ aids the charging capabilities of the battery, both faster and with higher maximum power. As will become clear in what follows, similar conditions are obtained for the \emph{collective} quantum battery

\subsection{Large-spin representations of the collective quantum battery-charger \label{subsec:large_spin}}

Among the many possibilities for the set of variable couplings that some NISQ platforms offer, we restrict our analysis to the situation where the onsite potentials of the battery and charger parts are taken as uniform, i.e. all battery's (charger's) qubits possess the same onsite potential $V_B$ ($V_C$). Furthermore, we set all couplings $J_{mn}$ of the qubit pairs within the battery (charger) in Eq.~\eqref{eq:HI} equal to $J_B$ ($J_C$), whereas all couplings between the battery and charger qubits equal to $J_{BC}$.

Within this prescription of homogeneous couplings for each part of the system, it is then convenient to define large spin operators of the battery and charger parts as
\begin{eqnarray}
S_B^{\alpha} = \sum_{m\in N_B} \frac{1}{2}\sigma_m^{\alpha},\quad
S_C^{\alpha} = \sum_{m\in N_C} \frac{1}{2}\sigma_m^{\alpha},
\label{eq:large-spin}
\end{eqnarray}
where $\alpha=x,y,z$. 
With these, the different Hamiltonian terms are simplified to
\begin{equation}
\begin{aligned}
    H_B &= V_B\left(S_B^z+\frac{N_B}{2}\right),\\
    H_C &= V_C\left(S_C^z+\frac{N_C}{2}\right),\\
    H_I &= J_{BC}\left(S_B^+S_C^-+S_B^-S_C^+\right)+\frac{J_B}{2}\left(S_B^+S_B^-+S_B^-S_B^+\right)\\
    &+\frac{J_C}{2}\left(S_C^+S_C^-+S_C^-S_C^+\right)-\frac{J_BN_B+J_CN_C}{2},
\end{aligned}
\label{eq:large_spinH}
\end{equation}
where the corresponding large-spin raising and lowering operators are
\begin{equation}
S_B^{\pm} = S_B^x \pm {\rm i} S_B^y,\quad
S_C^{\pm} = S_C^x \pm {\rm i}S_C^y.
\label{large-spin-raising}
\end{equation}
Since the $z$ component of the total spin $S^z=S_B^z+S_C^z$ is conserved, one can take ${\left|S_B^{z},S_C^{z}\right>}$ as basis and thus the initial state of Eq.~\eqref{eq:initial} is written as $\left|\psi (0)\right>=\left|\frac{N_B}{2},-\frac{N_B}{2}\right>_B\bigotimes \left|\frac{N_C}{2},\frac{N_C}{2}\right>_C$. 
\subsection{Low energy behaviors of the collective quantum battery-charger \label{subsec:low_energy}}

Based on the large-spin representations and the form of the initial state, an antiferromagnetic (AFM) Holstein-Primakoff (HP) transformation can be conveniently applied to each term in Eq.~\eqref{eq:large_spinH}. For the battery part, for example, we take 
$\left|\frac{N_B}{2},-\frac{N_B}{2}\right>=\left|\Downarrow\right>$ as the reference state such that the transformation reads 
\begin{equation}
\begin{aligned}
    S_B^z&=a^{\dagger}a-\frac{N_B}{2},\\
    S_B^{+}&=a^{\dagger}\sqrt{N_B-a^{\dagger}a},\\
    S_B^{-}&=\sqrt{N_B-a^{\dagger}a}a,
\end{aligned}
\label{batteryHP}
\end{equation}
where $a$($a^{\dagger}$) is the bosonic annihilation (creation) operator of the battery state. For the charger, the reference state is
$\left|\frac{N_C}{2},\frac{N_C}{2}\right>=\left|\Uparrow\right>$, and the corresponding transformation is written as 
\begin{equation}
\begin{aligned}
    S_c^z&=-b^{\dagger}b+\frac{N_C}{2},\\
    S_C^{+}&=\sqrt{N_C-b^{\dagger}b}b,\\
    S_C^{-}&=b^{\dagger}\sqrt{N_C-b^{\dagger}b},
\end{aligned}
\label{chargerHP}
\end{equation}
where, similarly, $b$($b^{\dagger}$) denotes the bosonic annihilation (creation) operator of the charger state. If we further assume that the battery is at the low energy state, and that excitations are small, $\left<a^{\dagger}a\right>\ll N_B$ and $\left<b^{\dagger}b\right>\ll N_C$, we obtain analytical expressions to describe the low-energy behavior of our quantum battery-charger, such that the Hamiltonian is written as
\begin{equation}
\begin{aligned}
    H_B&=V_Ba^{\dagger}a,\\
    H_C&=V_C(N_C-b^{\dagger}b),\\
    H_I&=J_{BC}\sqrt{N_BN_C}(a^{\dagger}b^{\dagger}+ab)+J_BN_Ba^{\dagger}a+\\
    &J_CN_Cb^{\dagger}b.
\end{aligned}
\label{HamiltonianHP}
\end{equation}
With the form of our initial state, the number of excitations is conserved, i.e., $\left<\psi(t)|D|\psi(t)\right>=0$, where $D=a^{\dagger}a-b^{\dagger}b$, resulting in  $\left<\psi(t)|a^{\dagger}a|\psi(t)\right>=\left<\psi(t)|b^{\dagger}b|\psi(t)\right>$. Finally, the total Hamiltonian $H$ reads 
\begin{equation}
H=\omega a^{\dagger}a+g(a^{\dagger}b^{\dagger}+ab)+V_CN_C,
\label{simpleHamiltonianH}
\end{equation}
where $\omega=J_BN_B+J_CN_C+\Delta V$ and $g=J_{BC}\sqrt{N_BN_C}$. Here and afterwards $\Delta V\equiv V_B-V_C$.
Given the form of the initial state, the time dependence of the battery's energy is
\begin{equation}
E_B(t)=\left\{
\begin{array}{lcl}
\frac{2g^2V_B}{\omega^2-4g^2}\left[1-\cos(\sqrt{\omega^2-4g^2}t)\right]  &   &\omega^2>4g^2 \\
V_Bg^2t^2    &   &\omega^2=4g^2 \\
\frac{2g^2V_B}{4g^2-\omega^2}\left[\cosh(\sqrt{4g^2-\omega^2}t)-1\right] &   &\omega^2<4g^2 \ .
\label{EB}
\end{array}    
\right.
\end{equation}
Its maximum is obtained under the condition $\omega^2>4g^2$:
\begin{equation}
E_B^{\rm max}=\frac{4g^2V_B}{\omega^2-4g^2},\quad t_E^{\rm max}=\frac{\pi}{\sqrt{\omega^2-4g^2}}, 
\end{equation}
where, as before, $t_E^{\rm max}$ gives the shortest time to reach $E_B^{\rm max}$. The corresponding average power of the battery reads
\begin{equation}
P_B(t)=\left\{
\begin{array}{lcl}
\frac{2g^2V_B}{(\omega^2-4g^2)t}\left[1-\cos(\sqrt{\omega^2-4g^2}t)\right]  &   &\omega^2>4g^2 \\
V_Bg^2t    &   &\omega^2=4g^2 \\
\frac{2g^2V_B}{(4g^2-\omega^2)t}\left[\cosh(\sqrt{4g^2-\omega^2}t)-1\right] &   &\omega^2<4g^2  \  ,
\label{PB}
\end{array}    
\right.
\end{equation}
whose maximum, again for $\omega^2>4g^2$, is written as
\begin{equation}
P_B^{\rm max}\simeq\frac{1.44g^2V_B}{\sqrt{\omega^2-4g^2}},\quad t_P^{\rm max}\simeq\frac{2.33}{\sqrt{\omega^2-4g^2}}.
\end{equation} 
Details of the calculations of Eqs.~\eqref{EB} and \eqref{PB} are included in Appendix \ref{sec:EP_B_AFM_HP}, where we further contrast the results stemming from the AFM-HP transformation with the ones obtained via exact diagonalization of the original Hamiltonian: A good match is observed at short times scales.

The expressions listed above for the case where $\omega^2>4g^2$ show that $E_B^{\rm max}$, $P_B^{\rm max}$, $t_E^{\rm max}$ and $t_P^{\rm max}$ have similar functional form to the corresponding ones for the parallel battery. Yet, as we shall see below, the differences in the parameters, in particular in the numbers of elements $N_B$ and $N_C$ yield significant differences in its extensive behavior. For example, if $N_B=N_C$ and $J_B=J_C=J_{BC}=J$ it results that $\omega^2-4g^2=4JN_B\Delta V+\Delta V^2$. If we further assume that $\Delta V>0$, and that $N_B$ is sufficiently large such that $4JN_B\gg \Delta V$, we end up with $\omega^2-4g^2\simeq4JN_B\Delta V$ and the physical quantities read
\begin{equation}
\begin{aligned}
E_B^{\rm max}&\simeq\frac{JN_BV_B}{\Delta V},\quad t_{E}^{\rm max}\simeq\frac{\pi}{\sqrt{4JN_B\Delta V}},\\
P_B^{\rm max}&\simeq\frac{0.72(JN_B)^{3/2}V_B}{\sqrt{\Delta V}},\quad t_{P}^{\rm max}\simeq\frac{2.33}{\sqrt{4JN_B\Delta V}}.   
\end{aligned}
\end{equation}
Here $\frac{J}{\Delta V}\ll1$ is assumed to guarantee that the low energy approximation is still valid. Thus, its dependence on the number of elements of the battery follows:
\begin{eqnarray}
E_B^{\rm max}\propto N_B;\  P_B^{\rm max}\propto N_B^{3/2};\  t_{E}^{\rm max}, t_{P}^{\rm max} \propto N_B^{-1/2}.
\label{eq:scaling}
\end{eqnarray}
One can see that $P_B^{\rm max}$ scales super-extensively with $N_B$, in stark contrast with the extensive scaling behavior of $P_B^{\parallel, {\rm max}}$ of the parallel case. 

\tcr{On the other hand, if we take a negative $\Delta V$, again with $N_B=N_C$, and $J_B=J_C=J_{BC}=J$, the scaling with the battery size changes. Within the condition that $\omega^2>4g^2$, it requires that $4JN_B+\Delta V<0$. If further assuming that $4JN_B\ll-\Delta V$, we end up with $\omega^2-4g^2\simeq\Delta V^2$ and the maximum values of the physical quantities read
\begin{equation}
\begin{aligned}
    E_B^{\rm max}&\simeq\frac{4J^2N^2_BV_B}{\Delta V^2},\quad t_{E}^{\rm max}\simeq\frac{\pi}{|\Delta V|},\\
P_B^{\rm max}&\simeq\frac{1.44J^2N_B^2V_B}{|\Delta V|},\quad t_{P}^{\rm max}\simeq\frac{2.33}{|\Delta V|}.
\end{aligned}
\end{equation}
Thus, its dependence on the number of elements of the battery follows:
\begin{equation}
    E_B^{\rm max},P_B^{\rm max}\propto N_B^2; t_{E}^{\rm max},t_{P}^{\rm max}\propto N_B^0.
\end{equation}
One can see that $E_B^{\rm }$ and $P_B^{\rm max}$ both scale super-extensively with $N_B$, with a power that is even larger than the ones with $\Delta V >0$.}
These two cases exemplify the quantum advantage of a quantum battery, here in a battery-charger model. 

\tcr{Now returning to the cases} $\omega^2\leq4g^2$, $E_B(t)$ is proportional to $t$, diverging at long times. Since the previous analytical expressions and accompanying scaling forms are based on the low energy approximation, once $E_B(t)$ is sufficiently large, the analytical results are no longer valid, and we instead make use of exact diagonalization to numerically investigate the dynamical behaviors of the physical quantities.

Before moving to these results, we comment that a recent `quantum-battery' study by some of us has employed a similar large-spin representation but on a different model~\cite{T_2021_PRA_Guan}. There, a ferromagnetic (FM) H-P transformation is applied instead. We point out that the type of H-P transformation to employ is largely initial-state-dependent: For the initial state of Eq.~\eqref{eq:initial} a two-large-spin scheme where the battery large-spin points down and the charger large-spin points up (thus exhibiting an AFM structure) an AFM transformation gives better results in comparison to the FM H-P one when contrasting both to the exact results (see Appendix~\ref{sec:benchmark} \tcr{for a detailed discussion}).

\section{Exact Numerical Results \label{sec:results}}
Combining the large-spin representation mentioned in Sec.~\ref{subsec:large_spin} and the ED method, we can investigate substantially larger system sizes than existing NISQ devices' current capabilities. The goal is first to characterize the aspects that maximize the capacity of quantum batteries through the inspection of the dynamics of physical quantities, then complement it with an analysis of the dependence on the system size, including the relative size of battery and charger parts. For better comparisons of $E_B(t)$, $P_B(t)$, and their maximum under different system sizes, we chose $V_B$ as the unit of energy and set it to 1 without loss of generality. 

\begin{figure}[htbp]
    \begin{center}
    \includegraphics[width=1\columnwidth]{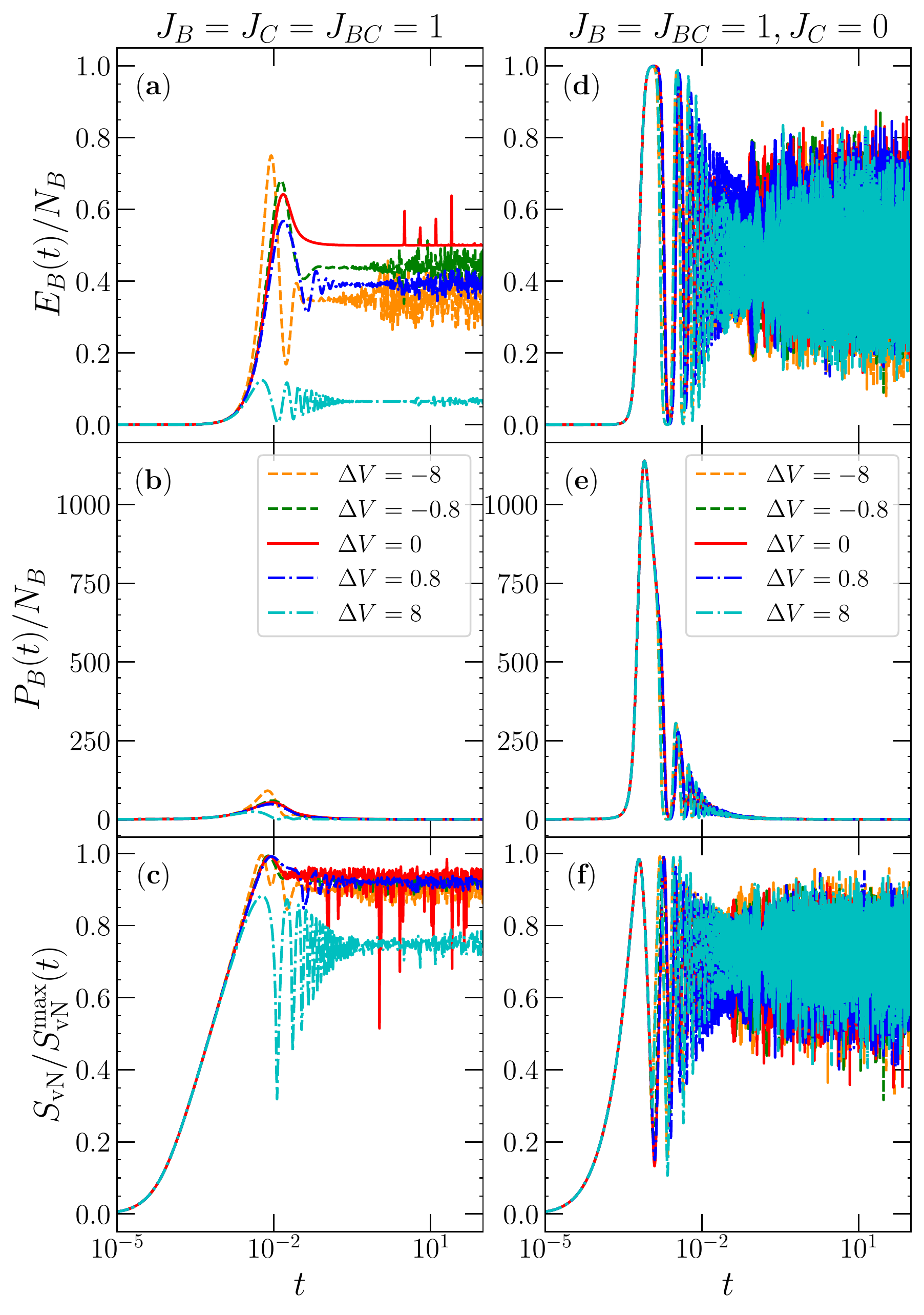}
    \end{center}
    \caption{Dynamics of various quantities for a large quantum battery-charger system featuring  $N_B=N_C=10,000$ qubits: The energy of the battery in (a)[(d)], the corresponding power in (b)[(e)], and lastly the von Neumann entropy (c)[(f)], describing the entanglement of the battery and charger parts. Left panels (a--c) describe the cases with homogeneous($J_B=J_C=J_{BC}=1$) couplings, whereas right panels (d--f) the same for the inhomogeneous couplings ($J_B=J_{BC}=1,J_C=0$). The difference in local energies of battery and charger is given by $\Delta V = V_B - V_C$ and we study five cases of $\Delta V$ with $V_B=1$.}
    \label{FIG2}
\end{figure}
\begin{figure}[htbp]
    \begin{center}
    \includegraphics[width=1.0\columnwidth]{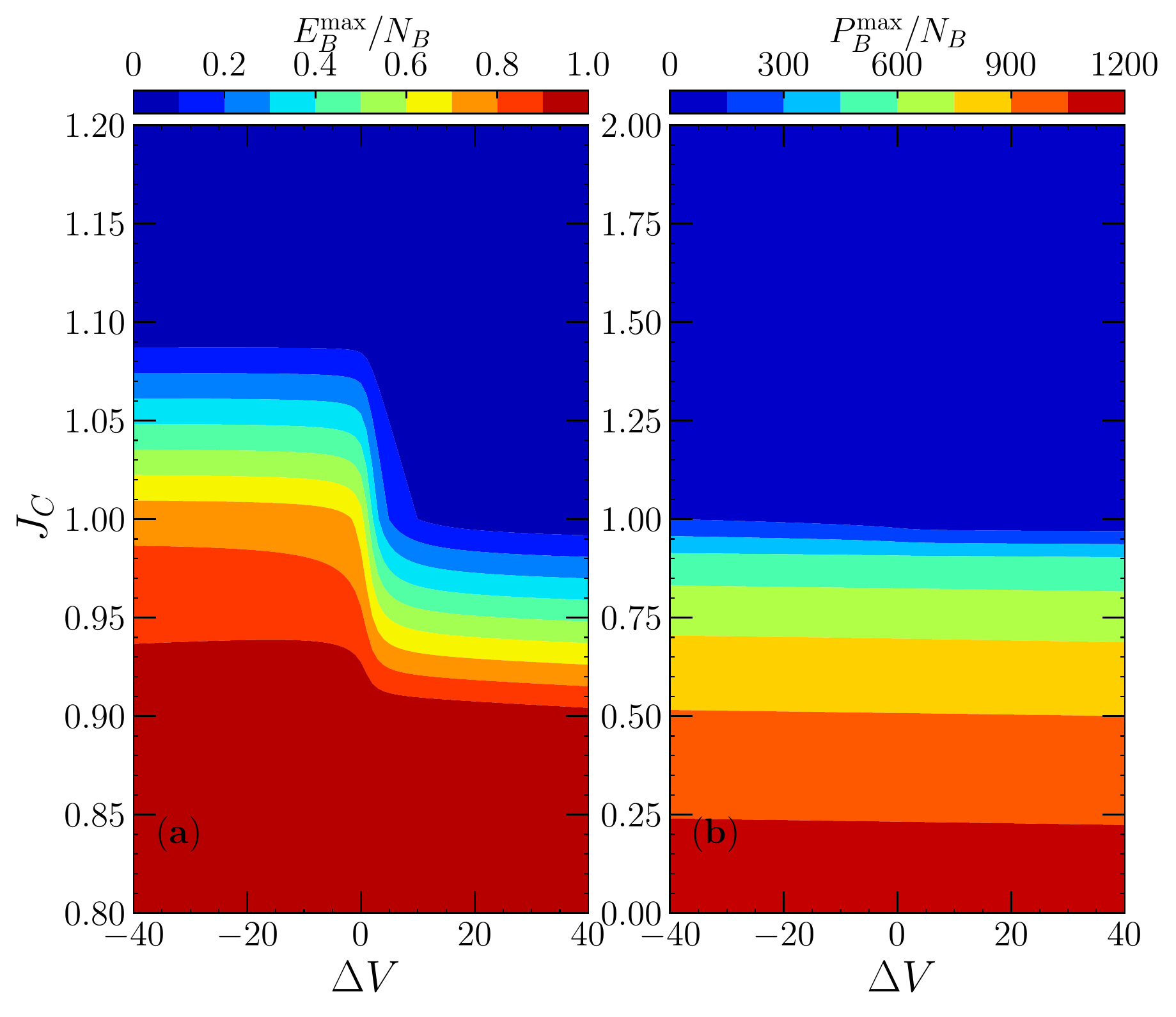}
    \end{center}
    \caption{\tcr{Contour plot of the (normalized by $N_B$) maximum energy (a) and power (b) in the battery as a function of the intra-charger coupling $J_C$ and the energy offset $\Delta V$. Here $N_B=N_C=10000$ and $J_B=J_{BC}=1$. Smaller magnitudes of $J_C$ decrease the connections among the charger qubits and help energy transfer; a negative $\Delta V$ implies a lower energy cost to transfer ``particles'' from the charger to the battery part which helps the energy transfer, too. These conclusions generalize the results of Fig.~\ref{FIG2}.}}
    \label{FIG2supplement}
\end{figure}
\subsection{Large quantum batteries: dynamics with $N_B=N_C$\label{subsec:evolution}}
 We report in Fig.~\ref{FIG2} the relaxation dynamics of $E_B(t)$, $P_B(t)$, accompanied by the corresponding  entanglement entropy $S_{\rm vN}$ [normalized by its maximum value $S^{\max}_{\rm vN}=\log(\min(N_B,N_C)+1)$], under different parameters of the total Hamiltonian. In particular, we contrast the cases of homogeneous($J_B=J_C=J_{BC}=1$) and inhomogeneous($J_B=J_{BC}=1,J_C=0$) couplings, with different choices of the onsite potentials $V_{B(C)}$, where $\Delta V = V_B-V_C$. The inhomogeneity from diminished $J_C$ is intuited by Ref.~\cite{T_2019_PRL_Andolina3} results and the idea that weakening the connections within the charger qubits aids energy transfer from the charger to the battery part.

\begin{figure*}[htbp]
    \begin{center}
    \includegraphics[width=0.75\textwidth]{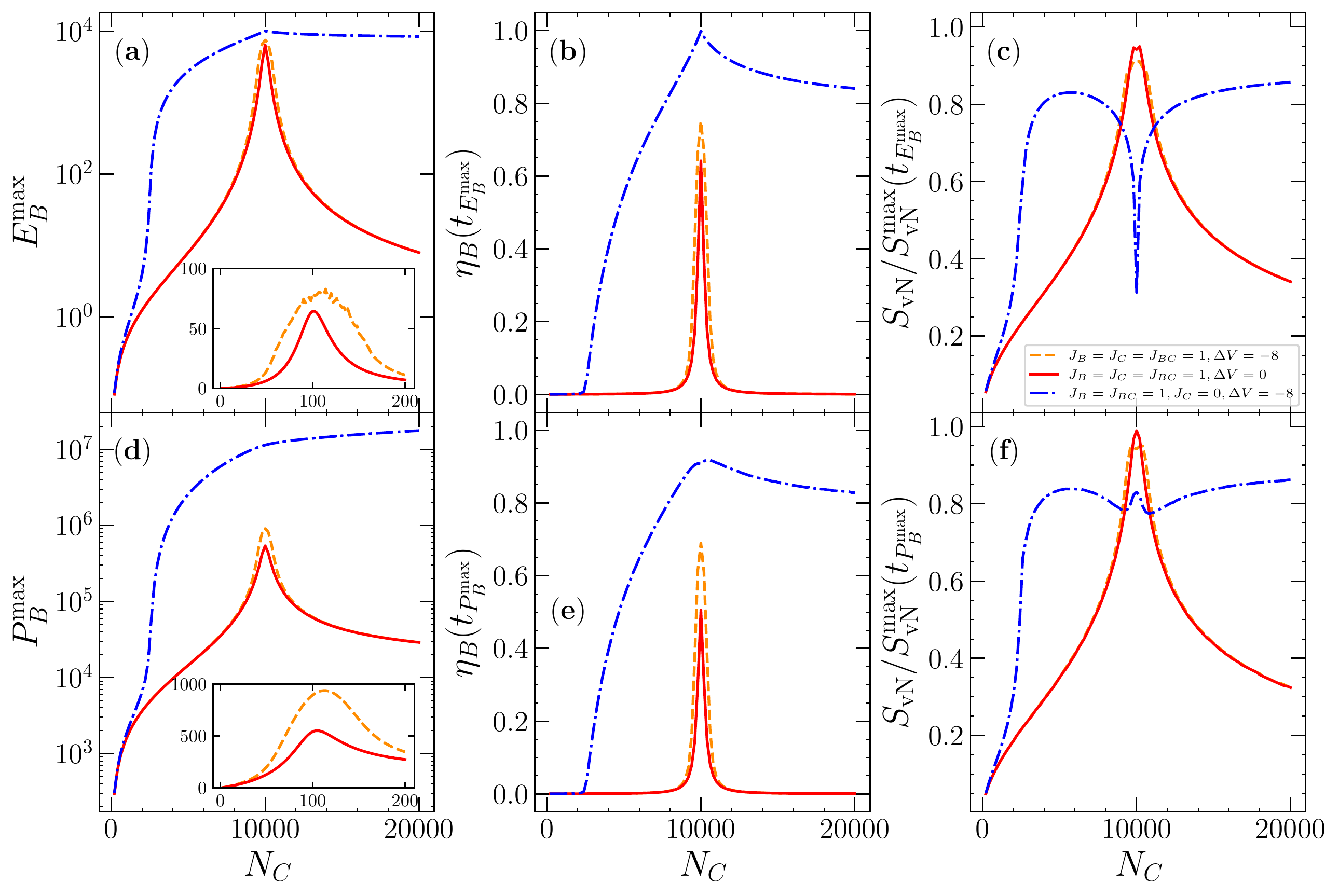}
    \end{center}
    \caption{(a) Maximum energy in the battery $E_B^{\rm max}$ and (d) maximum power $P_B^{\rm max}$ as a function of the number of charger qubits, for a battery with fixed number $N_B=10,000$. Insets give qualitatively similar results for a much smaller number of qubits (and potentially closer to existing NISQ devices), featuring $N_B=100$. The corresponding participation $\eta_B$ at the times where the maximum battery's energy (c) and power (e) are achieved. At those instants of time, (c) and (f) show the corresponding von Neumann entropy $S_{\rm vN}$(reduced by its maximum $S_{\rm vN}^{\rm max}$) of the battery part with the amount of the charger qubits $N_C$ under different couplings and onsite potentials. We contrast different situations, with either two homogeneous coupling cases ($J_B=J_C=J_{BC}=1$) or an inhomogeneous one ($J_B=J_{BC}=1,J_C=0$), with finite and vanishing energy offset between the charger and battery parts. Maximum energy transfer occurs when $N_B \simeq N_C$, despite variations among the different Hamiltonian parameters.}
    \label{FIG3}
\end{figure*}
Starting with the homogeneous coupling cases, we notice that the balanced situation with $\Delta V = 0$ leads to a significant build-up of $E_B$, with a quick equilibration reaching values close to half of the largest capacity of the battery part and subsequent revivals originated from integrability. On the other hand, if one introduces a negative offset on the energies of the battery-charger parts ($\Delta V = -8$ and $-0.8$), despite resulting in smaller $E_B$ at long times, it reaches a more significant maximum value at short times that surpasses the zero-offset case. \tcr{Such an enhancement is higher the larger the negative offset is.} Finally, in the opposite regime of positive energy offsets ($\Delta V$ = 8 and 0.8), energy flow is hampered, and the charger excitations are primarily contained throughout the dynamics. Furthermore, the larger the positive offset is, the stronger suppression it brings. 

\tcr{A straightforward interpretation is construed from these results:  In the bosonic language where the spin-operators map to, $V_B$ and $V_C$ represent the chemical potentials of the ``particles''  in the battery and charger parts. As such, the smaller $\Delta V$ is, the lower the energy cost required to transfer ``particles'' from the charger to the battery.} This suggests that tuning up the charger's onsite potentials aids energy transfer.

Such a tendency, however, is less obvious in inhomogeneous coupling cases. By setting $J_C=0$ we notice that $E_B/N_B$ exhibits large fluctuations around 0.5 at long times for different sets of $\Delta V$. At the same time, the short-time dynamics are similar irrespective of the value of the potential offset. However, the magnitudes of the maxima are larger and the time to reach them is shorter than those of homogeneous cases. Such enhancement is also observable in $P_B$[Fig.~\ref{FIG2}(e)]. It shows that decreasing the magnitude of the couplings within the charger \tcr{suppresses the energy transfer among the charger qubits and boosts that from the charger to the battery part. Even if $J_C$ is sufficiently small, the previously inferred drawbacks of the positive offsets of the onsite potentials are largely overcome, making the differences of onsite potentials rather irrelevant for the $\Delta V$'s investigated.}

\tcr{Similar dynamical features are also imprinted in the entanglement between the battery and charger parts shown in Figs.~\ref{FIG2}(c) and~\ref{FIG2}(f). For  homogeneous cases, the suppression of energy transfer results in a reduction of the von Neumann entropy for large, positive $\Delta V$.} Large fluctuations are also observed in the entanglement dynamics of inhomogeneous cases in Figs.~\ref{FIG2}(f) \tcr{at long times, directly reflecting the dynamics of the previous physical quantities.}

\tcr{Figure~\ref{FIG2supplement} compiles and summarizes the effects of enhancement of the maximum energy and power when changing both the energy offset $\Delta V$ and the intra-charger coupling $J_C$. The maximum energy in the battery [Fig.~\ref{FIG2supplement}] exhibits a sudden increase when $J_C$ is compatible with $J_B$ and $J_{BC}$ by turning $\Delta V$ from positive to negative. Such advantage of the negative $\Delta V$ gradually disappears as $J_C$ goes away from 1 where the magnitude of $E_B^{\rm max}$ is largely insensitive to the change of $\Delta V$. When $J_C$ is sufficiently large,  connections within the charger qubits are much stronger than those of the other parts, and energy preferentially flows within the charger. On the other hand, when $J_C$ is otherwise small, the initial energy of the charger can only be transferred to the battery part (even if only at short-time scales), which makes the sign of the offset irrelevant, too. The magnitude of $P_B^{\rm max}$ at $J_C \lesssim 1$ is significantly different to that of large $J_C$, shown in Fig.~\ref{FIG2supplement}(b). Especially when $J_C$ is smaller than 1, $P_B^{\rm max}$ slightly diminishes as $\Delta V$ is tuned from negative to positive, which implies the minor detrimental influences of positive offset.}

\tcr{Based on the conclusions for the different regime of parameters above, we pick out three representative configurations of the Hamiltonian and investigate how maximal quantities and entropy change with the system size in the next subsection.}

\begin{figure*}[htbp]
    \begin{center}
    \includegraphics[width=0.75\textwidth]{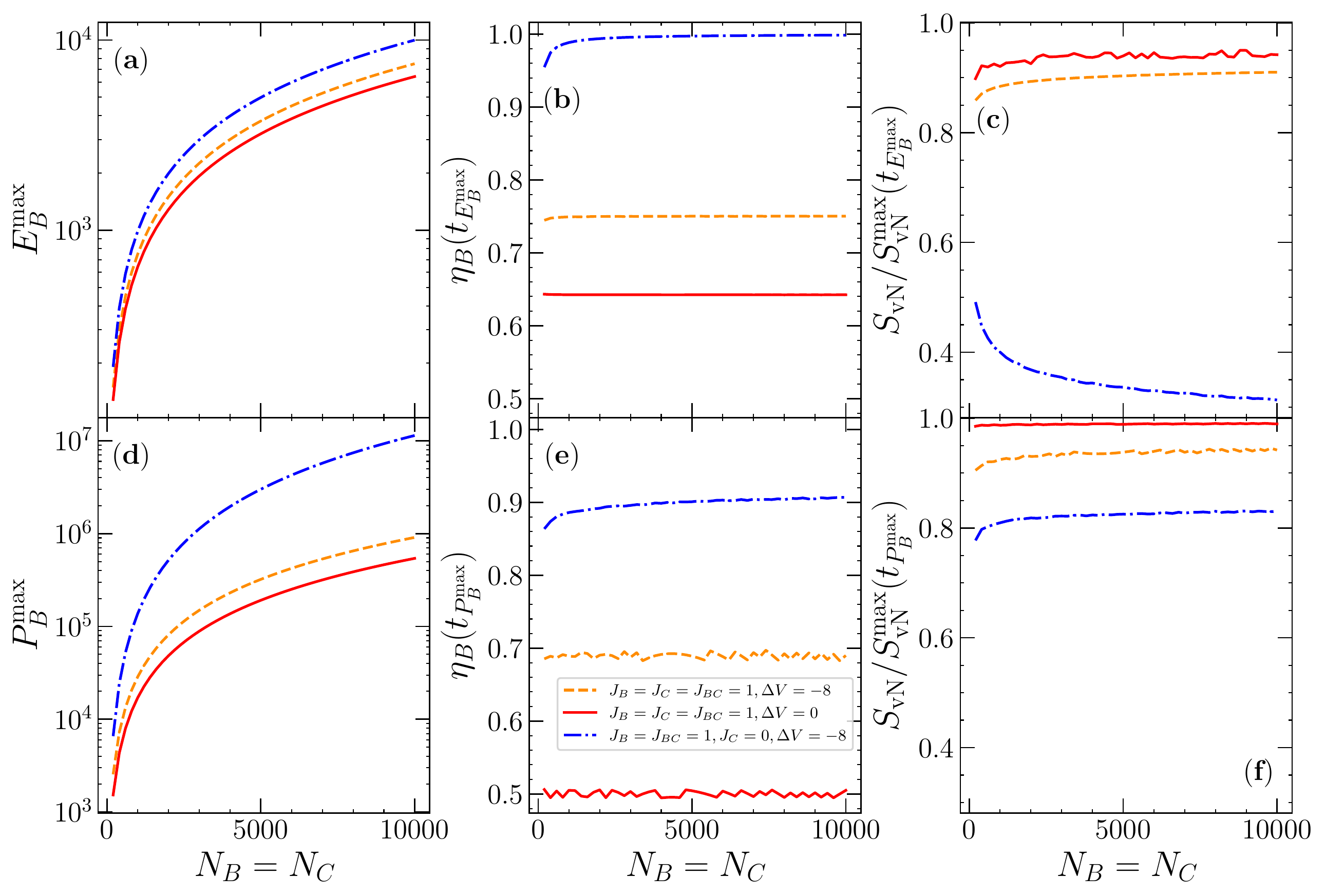}
    \end{center}
    \caption{Similar to Fig.~\ref{FIG3}, but now setting both battery and charger parts with the same size, i.e., $N_B=N_C$. The different Hamiltonian configurations, the same as the ones chosen in Fig.~\ref{FIG3}, lead to similar qualitative behavior, but energy and power in the battery are maximized in the case where the couplings have inhomogeneity, ($J_B=J_{BC}=1,J_C=0$), and a negative onsite energy offset($\Delta V = -8$) between both parts.}
    \label{FIG4}
\end{figure*}

\subsection{Imbalance of quantum battery-chargers:  $N_B\neq N_C$ \label{sec:NB_neq_NC}}
\indent Besides the energy and power of the battery [Eq.~\eqref{eq:E_B_P_B}], we further characterize the `participation' of the battery part, 
\begin{equation}
\eta_B(t)=\frac{\langle\psi(t)|\sum_{m\in N_B}\sigma_m^+\sigma_m^-|\psi(t)\rangle}{\min(N_B,N_C)}.
\label{eq:participation}
\end{equation} 
It physically represents the ratio of the acquirable `particles' the battery has obtained over the course of the dynamics. Having established general features that maximize the energy charger-battery transfer for systems with the same number of qubits in each part, we now investigate how the various quantities are affected once we allow that $N_B\neq N_C$. 

First, we analyze the case where the battery size is fixed ($N_B = 10,000$), while changing the number of charger's qubits from 200 to 20,000. Results for this situation are summarized in Fig.~\ref{FIG3}. To start, we show $E_B^{\rm max}$ and $P_B^{\rm max}$ vs.~the charger qubit number $N_C$ in Figs.~\ref{FIG3}(a) and \ref{FIG3}(d): For homogeneous couplings a `resonant' condition is achieved, i.e., both energy and power in the battery are maximized once $N_B \simeq N_C$. Inhomogeneity brought by a suppressed $J_C$ enhances the energy transfer and $E_B^{\rm max}$ and $P_B^{\rm max}$ are larger than those of the homogeneous cases. When $N_C>N_B$, the maxima do not experience a steep decay. Such enhancement of energy transfer originating from inhomogeneity is similar to what we observe in Fig.~\ref{FIG2} with $N_B=N_C$. Whereas those qubit numbers are much beyond what present-day NISQ platforms can coherently emulate, we note that fixing $N_B=100$ and changing the number of charger qubits up to $N_C=2N_B$ leads to qualitatively similar results, as shown in the insets in Figs.~\ref{FIG3}(a) and \ref{FIG3}(d).

Observing the participation $\eta_B$ at instants of time that maximize both the energy and power [Figs.~\ref{FIG3}(b) and \ref{FIG3}(e)], the rapid increase displayed in $E_B^{\rm max}$ and $P_B^{\rm max}$ of the homogeneous cases at $N_B \simeq N_C$ is also reflected on the corresponding $\eta_B$ value. Moreover, a reduced coupling ($J_C=0$) with finite energy offset ($\Delta V =-8$) shows that the `particle' occupation in the battery is largely enhanced, and the charger-battery energy flow is boosted by this inhomogeneity. Lastly, $S_{\rm vN}$ between the two parts of the system [Figs.~\ref{FIG3}(c) and \ref{FIG3}(f)] shows behavior reminiscent of the previous quantities: `Resonant' peaks appear, and furthermore considerable improvement of energy transfer makes the battery part charged close to its highest energy state, which is a pure state of the whole system with zero entanglement. Thus a decrease of von Neumann entropy of the inhomogeneous case in Figs.~\ref{FIG3}(c) (blue dashed-dotted lines) can be observed when $N_B\sim N_C$.

One aspect that has since been used to describe an improvement of the work extraction in quantum battery systems is the capping of the amount of entanglement in the course of the dynamics, accomplished in Ref.~\cite{T_2019_PRB_Polini} by the introduction of disorder on the onsite energy levels. This is related to the onset of a many-body localized phase, known to reduce entanglement in quantum systems away from equilibrium~\cite{Bardarson2012,Serbyn2013}. Here in our case, the barrier to entanglement build-up within the charger was accomplished by the reduced coupling in this part of the system. The previously described `resonant' condition for energy transfer is thus significantly modified, and now a wide range of $N_C$ values can already be seen to improve the energy transfer between charger and battery [Fig.~\ref{FIG3}(a)].

Building on those results, we now investigate a second configuration, one with an equal number of battery and charger qubits, $N_B=N_C$, while monitoring the influence of the total system size in Fig.~\ref{FIG4}. As one would expect, $E_B^{\rm max}$ [Fig.~\ref{FIG4}(a)] and $P_B^{\rm max}$[Fig.~\ref{FIG4}(d)] steadily grow with the system, but the situation with inhomogeneous couplings ($J_B=J_{BC}=1,J_C=0$) and a finite offset energy between charger and battery ($\Delta V = -8$) displays the largest overall increase. This occurs with the largest participations $\eta_B$ [Figs.~\ref{FIG4}(b) and \ref{FIG4}(e)] among the different Hamiltonian parameters used. Yet, this regime is highlighted by a small entanglement entropy due to the approach to the highest energy state of the battery at the instants of time where the maxima energy and power are achieved in the battery [Figs.~\ref{FIG4}(c) and \ref{FIG4}(f)].

\subsection{Scaling forms}
\begin{figure}[htbp]
    \begin{center}
    \includegraphics[width=0.9\columnwidth]{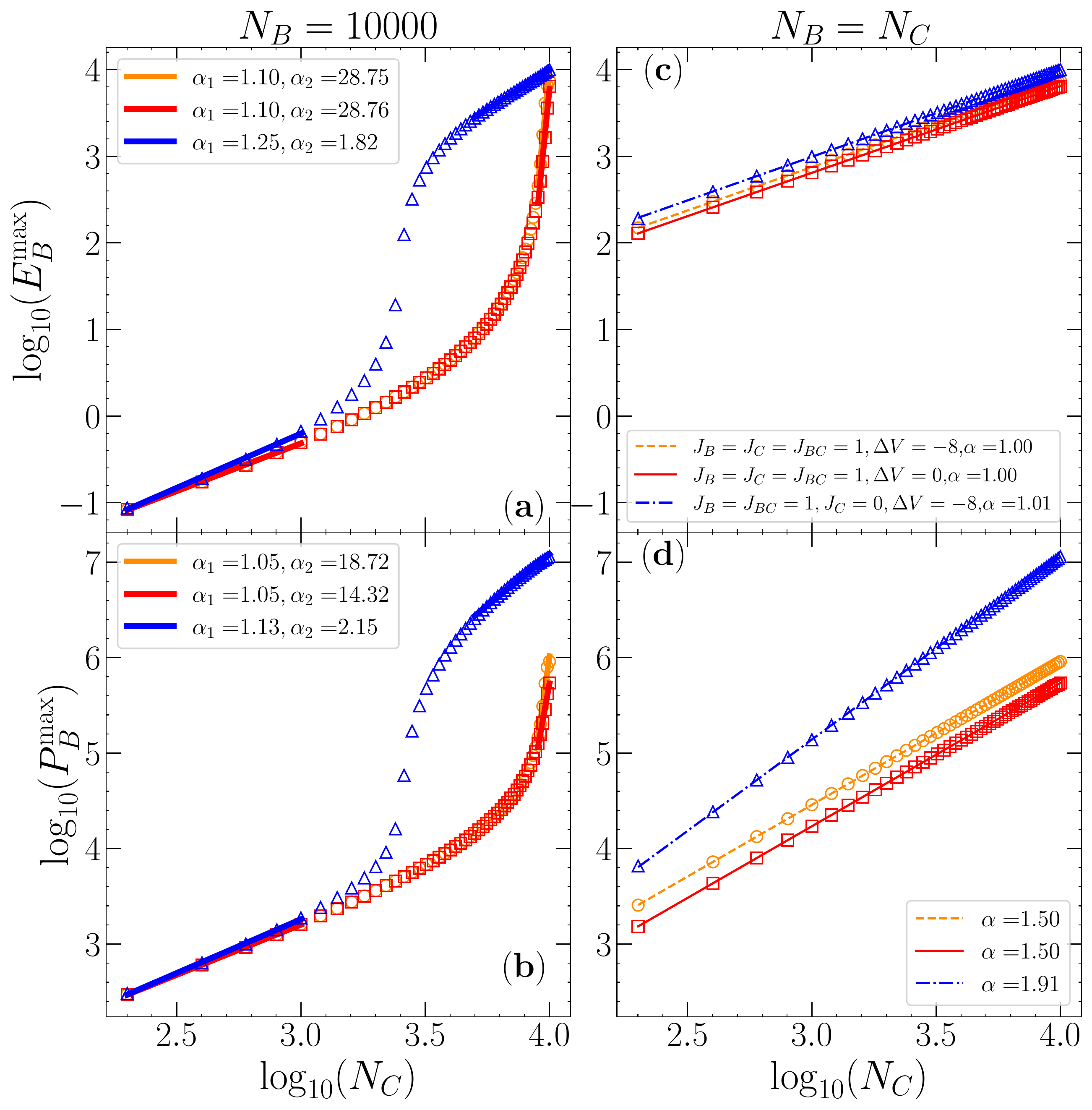}
    \end{center}
    \caption{Scaling of $E_B^{\rm max}$ (upper panels) and $P_B^{\rm max}$ (lower panels) with $N_C^{\alpha}$, for two settings of collective quantum batteries: (a, b) with fixed battery size ($N_B = 10,000$ -- see Sec.~\ref{subsec:evolution}), and (c, d) with equal amount of qubits in each part, $N_B=N_C$ (see Sec.~\ref{sec:NB_neq_NC}). The latter displays superextensive scalings, with $\alpha \geq 1.5$. The fixed $N_B$ case, on the other hand, is marked by different scaling forms, depending on whether $N_B$ is close to $N_C$ or not. Lines depict linear fittings on the log-log plot, for the data within the corresponding range they are plotted.}
    \label{FIG5}
\end{figure}

With the overall dependence of $E_B^{\rm max}$ and $P_B^{\rm max}$ in the battery size established, we turn now to a specific quantification of the growth with the charger size. In the regime where the number of qubits within the battery and charger parts are evenly matched, as originally shown in Fig.~\ref{FIG4}, the maximal energy and power grow as a power-law with $N_C$, i.e., $E_B^{\rm max},\ P_B^{\rm max}  \propto (N_C)^\alpha$. As Fig.~\ref{FIG5}(c) shows, irrespective of the different Hamiltonian configurations, the growth of the maximal energy is consistent with an extensive behavior, $\alpha \simeq 1$. More surprising scaling, however, can be observed in the maximal power shown in \ref{FIG5}(d). Here, the scaling exponent $\alpha$ of the homogeneous cases is approximately 1.5, which shows that a `quantum advantage' is present in such settings. What is more, the decrease of $J_C$ strengthens this superextensive scaling exponent to values close to 1.9. In this regime, one is close to saturating the more stringent bounds on the quantum advantage recently demonstrated under collective charging~\cite{Gyhm2022}.

A further check of $t_{P_B^{\rm max}}$, the corresponding time to reach $P_B^{\rm max}$ over such settings, shown in Fig.~\ref{FIG9}(b) in Appendix \ref{sec:t_Pmax}, confirms that the superextensive behavior of $P_B^{\rm max}$ originates from the subextensive nature of $t_{P_B^{\rm max}}$, while the energy scaling is always extensive. Whereas the ideal conditions that led to these results can be interpreted as non-realistic for experimental emulation, the superextensive scaling of $P_B^{\rm max}$ and subextensive behavior of $t_{P_B^{\rm max}}$ are yet observed even when the couplings acquire some noise, shown in Fig.~\ref{FIG10}. This confirms the robustness of the claimed superextensive scaling in Fig.~\ref{FIG5}(d), for not being tied to a specific ideal parameter setting.

At the same time, a direct comparison to the cases where the battery size is fixed ($N_B = 10,000$) while one changes the number of qubits in the charger $N_C$, is shown in Figs.~\ref{FIG5}(a) and \ref{FIG5}(b). Here one can classify the dependence of either $E_B^{\rm max}$ or $P_B^{\rm max}$ with $N_C$ via two types of power law, $(N_C)^{\alpha_1}$ when $N_B \ll N_C$, and $(N_C)^{\alpha_2}$ when $N_B\to N_C$. While in the former, one can immediately see that $1 < \alpha_1 \lesssim 2$, the scaling when approaching the previously described `resonant' condition is much more robust. In concrete terms, for the cases of the homogeneous couplings, $\alpha_2$ for the considered range of $N_C$ charger qubits are larger than 25, with either a finite offset $\Delta V$ or with $\Delta V = 0$. Investigation of $t_{P_B^{\rm max}}$ of the $N_B$-fixed cases depicted in Fig.~\ref{FIG9}(a) expresses similar `resonant' peaks when $N_B\sim N_C$ in the homogeneous coupling cases. For the inhomogeneous case, a peak appears when $N_C$ is much smaller to $N_B$ instead.

\section{Conclusions}\label{sec:conclusions}
In summary, we have investigated a quantum battery-charger model, composed of interacting qubits, with a specific focus on existing NISQ platforms. First, by introducing a large-spin representation to simplify the Hamiltonian under the assumption of homogeneous couplings and onsite potentials, we extracted analytical forms of the battery's energy $E_B(t)$, power $P_B(t)$ (as well as their maximum values $E_B^{\rm max}$, $P_B^{\rm max}$) based on the AFM H-P transformation and low energy approximation. Within these conditions, we show the extensive and superextensive scalings of $E_B^{\rm max}$ and $P_B^{\rm max}$, respectively. Combining the large-spin representation and ED, we numerically investigate the real-time dynamics of $E_B(t)$, $P_B(t)$, participation $\eta_B$ and entanglement, represented by $S_{\rm vN}$, between the battery and charger part, revealing the set of Hamiltonian parameters that maximize the energy transferred to the battery part. 

Finally by checking the scalings of $E_B^{\rm max}$, $P_B^{\rm max}$ with $N_C$, the number of charger qubits under two settings, for either $N_B$ fixed or $N_B=N_C$, we observe a `resonant' condition in the former case that maximizes the energy transferred, in particular in the homogeneous-coupling cases, accompanied by close relations of the quantum entanglement. Inhomogeneity brought by a reduced $J_C$ leads to a performance that surpasses that of the homogeneous cases in energy transfer. Similar results have been obtained in Ref.~\cite{T_2019_PRL_Andolina3}, in that connections among charger elements suppress energy transfer from this part. Superextensive scaling of $P_B^{\rm max}$, interpreted as a signal of quantum advantage, can be observed with the growth of the system size when setting $N_B=N_C$, which is robust even if the couplings of qubits have an included noise. 

Superextensive behavior observed in physical quantities, especially the maximal power is treated as evidence of quantum advantages of quantum batteries against their classical counterparts, and has been reported in many studies~\cite{T_2017_PRA_Felix,T_2018_PRL_Marco,T_2018_AXRIV_blaauboer}. Other investigations, however, point out that after properly characterizing the thermodynamic limit some of the then-claimed superextensive behaviors disappear (see Sec.~\ref{sec:introduction}). The idea of carefully characterizing the thermodynamic limit was exposed in Ref.~\cite{E_2017_PRL_Kavan}, to avoid the unfairness of comparisons between the parallel and collective batteries, the latter introducing extra energy of battery elements by their interactions to drive transitions. In this work, the battery-charger system is isolated from the environment, and, most importantly, due to the special choice of initial (product) state, additional interactions of the collective battery over the parallel case just provide ``collective effects'' with no extra energy introduced once the onsite potentials are initially fixed. This is the reason why the maximal energy in the battery scales extensively when $N_B=N_C$. Thus the parallel and collective batteries can be fairly compared by setting the same system sizes, under the same conditions for onsite potentials and initial state we choose.

While many concrete aspects of quantum batteries remain elusive for a precise technical application, as efficient means to extract the stored energy after the charging process, for example, our investigation provides a path and quantitative aspects for their emulation in platforms featuring superconducting qubits, departing from abstract models that merely introduce chargers as effective external fields. Under our quantum battery-charger platform, our results support the idea of reducing the entanglement within the charger, by mitigating their coupling, as a fundamental step in achieving an efficient charging process, or that a negative battery-charger energy offset is preferable to increase power within a given system size where couplings are homogeneous.

\noindent
\begin{acknowledgments}
We thank Qiujiang Guo for insightful discussions. We acknowledge financial support from National Natural Science Foundation of China (NSFC) Grants No. NSAF U1930402, 12088101, 12050410263, 12111530010 and 11974039. Computations were performed on the Tianhe-2JK at the Beijing Computational Science Research Center.  
\end{acknowledgments}

\appendix

\section{Details of the parallel battery calculations \label{app:parallel}}
\indent For completeness, we introduce the details of the parallel-charging protocol, schematically displayed in Fig.~\ref{FIG1}(a). To start, in consistency to the case of an all-to-all coupled quantum battery, we set the initial state of each pair as $|\psi(0)\rangle^{\rm pair}=|\downarrow_b\uparrow_c\rangle$, in which the battery part is `empty' while the charger is `full'. Since the total spin $s=s_b+s_c$ of the pair is conserved, we can choose $\{\left|\downarrow_b\uparrow_c\right>,\left|\uparrow_b\downarrow_c\right>\}$ as the basis to write $H^{\rm pair}$ defined in Eq.~(\ref{eq:pair}) in the main text:
\begin{equation}
H^{\rm pair}=\left(
    \begin{matrix}
        V_c & J_{\rm pair}\\
        J_{\rm pair}  &   V_b
    \end{matrix}
    \label{appA:totalH}
\right)
\end{equation}
where the battery Hamiltonian is 
\begin{equation}
H_b^{\rm pair}=\left(
    \begin{matrix}
        0 & 0\\
        0  &   V_b
    \end{matrix}
\right)
\end{equation}
The corresponding eigenvalues of $H^{\rm pair}$ are:
\begin{equation}
    \lambda_1=\frac{V-\Omega}{2}\quad
    \lambda_2=\frac{V+\Omega}{2}
\end{equation}
with $V=V_b+V_c$, $\Delta V=V_b-V_c$ and $\Omega=\sqrt{4J_{\rm pair}^2+\Delta V^2}$. Its eigenvectors read
\begin{equation}
\begin{aligned}
    v_1&=\left(-\frac{\Delta V+\Omega}{2aJ_{\rm pair}},\frac{1}{a}\right)^\intercal\\
    v_2&=\left(-\frac{\Delta V-\Omega}{2bJ_{\rm pair}},\frac{1}{b}\right)^\intercal \ ,
\end{aligned}
\end{equation}
where $a^2=\frac{\Omega^2+\Delta V\cdot\Omega}{2J^2_{\rm pair}}$ and $b^2=\frac{\Omega^2-\Delta V\cdot\Omega}{2J^2_{\rm pair}}$ are normalizing parameters. The wavefuntion at time $t$ following an unitary evolution is written as:
\begin{equation}
|\psi(t)\rangle^{\rm pair}=
\begin{pmatrix}
    \frac{\Omega+\Delta V}{2\Omega}e^{-{\rm i}\lambda_1t}+\frac{\Omega-\Delta V}{2\Omega}e^{-{\rm i}\lambda_2t}\\
    -\frac{J_{\rm pair}}{\Omega}e^{-{\rm i}\lambda_1t}+\frac{J_{\rm pair}}{\Omega}e^{-{\rm i}\lambda_2t}
    \label{app:state_t}
\end{pmatrix}
\end{equation}
and thus the instantaneous energy $E_b^{\rm pair}=\langle\psi(t)|^{\rm pair}H_b^{\rm pair}|\psi(t)\rangle^{\rm pair}$ is finally written as
\begin{equation}
    E_b^{\rm pair}=\frac{2V_bJ_{\rm pair}^2}{\Omega^2}\left[1-\cos(\Omega t)\right]\ ,
\end{equation}
whose average power is
\begin{equation}
    P_b^{\rm pair}=\frac{2V_bJ_{\rm pair}^2}{\Omega^2t}\left[1-\cos(\Omega t)\right]
\end{equation}
The density matrix of the pair, $\rho^{\rm pair}=|\psi(t)\rangle^{\rm pair}\langle\psi(t)|^{\rm pair}$, is used to write entanglement entropy between the two qubits of the pair $S_{\rm vN}^{\rm pair}$:
\begin{equation}
    S_{\rm vN}^{\rm pair}=-A\log(A)-B\log(B)
\end{equation}
where $A=\cos^2(\frac{\Omega t}{2})+\frac{\Delta V^2}{\Omega^2}\sin^2(\frac{\Omega t}{2})$ and $B=\frac{4J_{\rm pair}^2}{\Omega^2}\sin^2(\frac{\Omega t}{2})$.
Then Eqs.~(\ref{eq:para-energy}-\ref{eq:para-von}) in the main text are demonstrated.

\begin{figure}[htbp]
    \begin{center}
    \includegraphics[width=0.70\columnwidth]{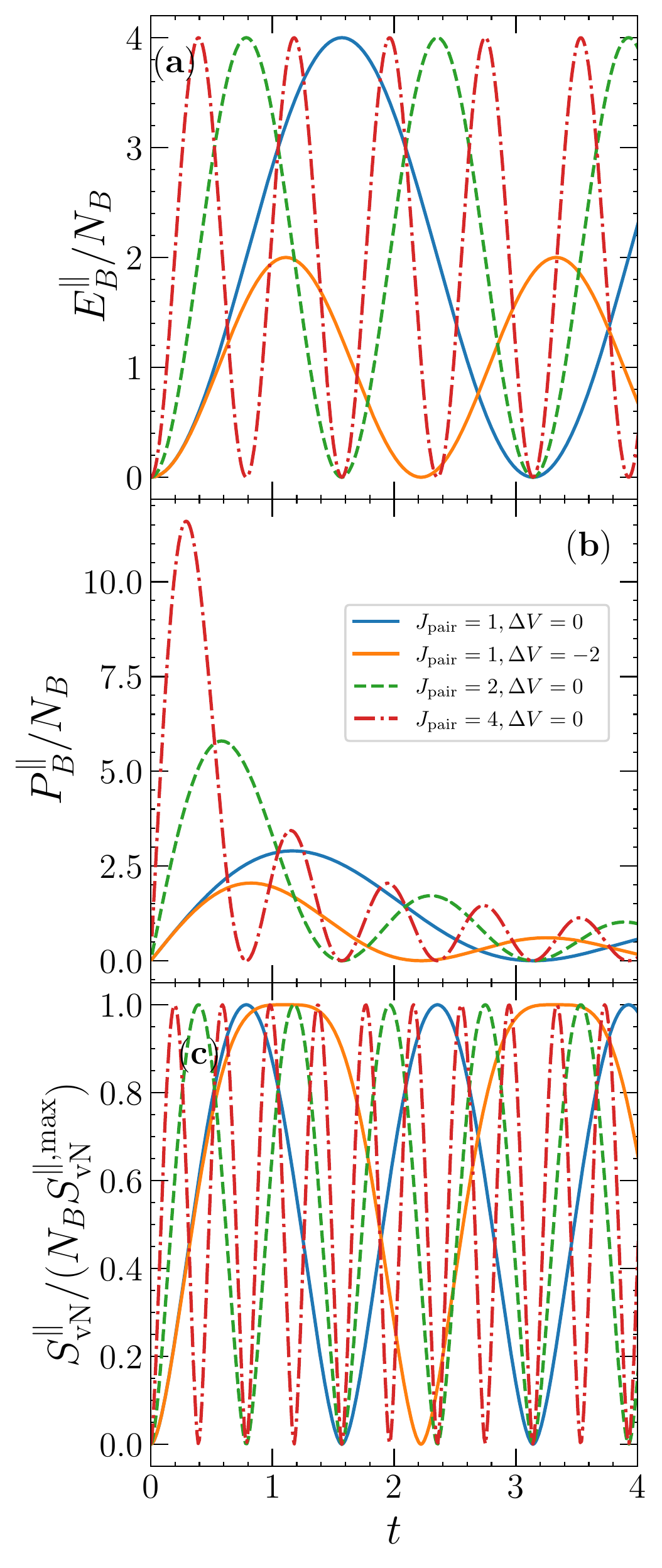}
    \end{center}
    \caption{Dynamics of $E_B^{\parallel}$ (a), $P_B^{\parallel}$ (b) and $S_{\rm vN}^{\parallel}$, normalized by the maximum $S_{\rm vN}^{\parallel,\rm max}$, (c) for different parameters of the parallel battery as shown in Fig.~\ref{FIG1}(a). The battery qubit cannot be fully charged if the detuning $\Delta V\neq0$. Improving the magnitude of $J_{\rm pair}$ trivially enhances $P_B^{\parallel, {\rm max}}$. }
    \label{FIG6}
\end{figure}

The dynamics of these three quantities is reported in Fig.~\ref{FIG6}, for various combinations of the battery-charger qubit coupling $J_{\rm pair}$ and the energy offset $\Delta V$. As expected, no steady state is observed in this effective two-qubit problem, but rather an oscillation characterized by $\Omega$ that grows with the coupling $J_{\rm pair}$ and the energy offset $\Delta V$. Additionally, a finite $\Delta V$ suppresses the charging, whereas increasing the magnitude $J_{\rm pair}$ trivially increases the maximum energy stored and the corresponding power.

Further characterization of the charging process is captured by the entanglement entropy, $S_{\rm vN}$ [see Eq.~\ref{eq:entropy}], depicted for the same set of parameters in Fig.~\ref{FIG6}(c). In the whole procedure, either in this parallel case or the collective one investigated in the details in the main text, the dynamics starts from a product state, whose associated entanglement entropy thus vanishes. Focusing on $\Delta V = 0$, the instants of time where the maximum energy is deposited in the battery corresponds to another product state, with excitations reversed in comparison to the initial preparation, $|\psi(t_{E_{B}^{\rm max}})\rangle^{\rm pair}=|\uparrow_b\downarrow_c\rangle$. As a result, $S_{\rm vN}$ also vanishes at $t = t_{E_{B}^{\rm max}}$. Conversely, at instants of time where $E_B^\parallel$ is half of its maximum charge, i.e., the energy is evenly distributed among charge and battery parts, the wavefunction is maximally entangled among its constituents and the entanglement entropy reaches a maximum. This is immediately observed in  Figs.~\ref{FIG6}(a) and (c), where the inflection points of the former lead to peaks in the latter.

As previously mentioned, a finite $\Delta V$ complicates this picture, since the battery is never maximally charged. Nonetheless, at times at which the system periodically returns to $E_B = 0$, $S_{\rm vN} = 0$.

\section{Analytic calculations of $E_B$ and $P_B$ of the Quantum Battery} \label{sec:EP_B_AFM_HP}
The concrete form of the energy of the battery reads
\begin{equation}
\begin{aligned}
    H_B&=V_B\left<\psi(t)|a^{\dagger}a|\psi(t)\right>\\
        &=V_B\left<\psi(0)|e^{iHt}a^{\dagger}ae^{-iHt}|\psi(0)\right>  \ .
\end{aligned}
\end{equation}
The key step is the calculation of $e^{{\rm i}Ht}a^{\dagger}ae^{-{\rm i}Ht}$, which based on the definition of Baker–Campbell–Hausdorff formula, one can obtain the following results (for the first 6 orders):
\begin{equation}
\begin{aligned}
1{\rm st}:    \left[H,a^{\dagger}a\right]&=g(-a^{\dagger}b^{\dagger}+ab)\\
2{\rm nd}:    \left[H,1{\rm st}\right]&=-g\omega(a^{\dagger}b^{\dagger}+ab)-2g^2(a^{\dagger}a+bb^{\dagger})\\
3{\rm rd}:    \left[H,2{\rm nd}\right]&=-g(\omega^2-4g^2)(a^{\dagger}b^{\dagger}-ab)\\
4{\rm th}:    \left[H,3{\rm rd}\right]&=-g\omega(\omega^2-4g^2)(a^{\dagger}b^{\dagger}+ab)\\
                        &-2g^2(\omega^2-4g^2)(a^{\dagger}a+bb^{\dagger})\\
5{\rm th}:    \left[H,4{\rm th}\right]&=-g(\omega^2-4g^2)^2(a^{\dagger}b^{\dagger}-ab)\\
6{\rm th}:    \left[H,5{\rm th}\right]&=-g\omega(\omega^2-4g^2)^2(a^{\dagger}b^{\dagger}+ab)\\
                        &-2g^2(\omega^2-4g^2)^2(a^{\dagger}a+bb^{\dagger})                        
\end{aligned}
\end{equation}
For our initial state $\left|\Downarrow_B,\Uparrow_C\right>$ only the even orders containing the term $a^{\dagger}a+bb^{\dagger}$ have non-vanishing contributions. Thus $E_B$ reads,
\begin{equation}
    E_B=2g^2V_B\left[\frac{t^2}{2}-\frac{t^4}{4!}(\omega^2-4g^2)+\frac{t^6}{6!}(\omega^2-4g^2)^2+\dots\right]\ .
\end{equation}
If $\omega^2=4g^2$, the battery's energy is 
\begin{equation}
E_B=V_Bg^2t^2\ ,
\end{equation}
and the corresponding power is written as
\begin{equation}
P_B=V_Bg^2t\ .
\end{equation}
In the case $\omega^2>4g^2$, $E_B$ simplifies to 
\begin{equation}
\begin{aligned}
    E_B&=\frac{2g^2V_B}{\omega^2-4g^2}\left[\frac{t^2}{2}(\sqrt{\omega^2-4g^2})^2-\frac{t^4}{4!}(\sqrt{\omega^2-4g^2})^4+\right.\\
    &\left.\frac{t^6}{6!}(\sqrt{\omega^2-4g^2})^6+\dots\right]\\
    &=\frac{2g^2V_B}{\omega^2-4g^2}\left[1-\cos(\sqrt{\omega^2-4g^2}t)\right]\\
\end{aligned}
\end{equation}
with $P_B$ reading as
\begin{equation}
P_B=\frac{2g^2V_B}{(\omega^2-4g^2)t}\left[1-\cos(\sqrt{\omega^2-4g^2}t)\right] \ .
\end{equation}
Whereas for $\omega^2<4g^2$, $E_B$ and $P_B$ read 
\begin{equation}
\begin{aligned}
    E_B&=\frac{2g^2V_B}{4g^2-\omega^2}\left[\frac{t^2}{2}(\sqrt{4g^2-\omega^2})^2+\frac{t^4}{4!}(\sqrt{4g^2-\omega^2})^4+\right.\\
    &\left.\frac{t^6}{6!}(\sqrt{4g^2-\omega^2})^6+\dots\right]\\
    &=\frac{2g^2V_B}{4g^2-\omega^2}\left[\cosh(\sqrt{4g^2-\omega^2}t)-1\right]\\
\end{aligned}
\end{equation}
and
\begin{equation}
P_B=\frac{2g^2V_B}{(4g^2-\omega^2)t}\left[\cosh(\sqrt{4g^2-\omega^2}t)-1\right]\ ,
\end{equation}
respectively. These correspond to the final forms of $E_B$ and $P_B$ shown in the main text, Eqs.~(\ref{EB}) and (\ref{PB}).

\section{Benchmarking the H-P transformation \label{sec:benchmark}}
\begin{figure}[htbp]
    \begin{center}
    \includegraphics[width=0.99\columnwidth]{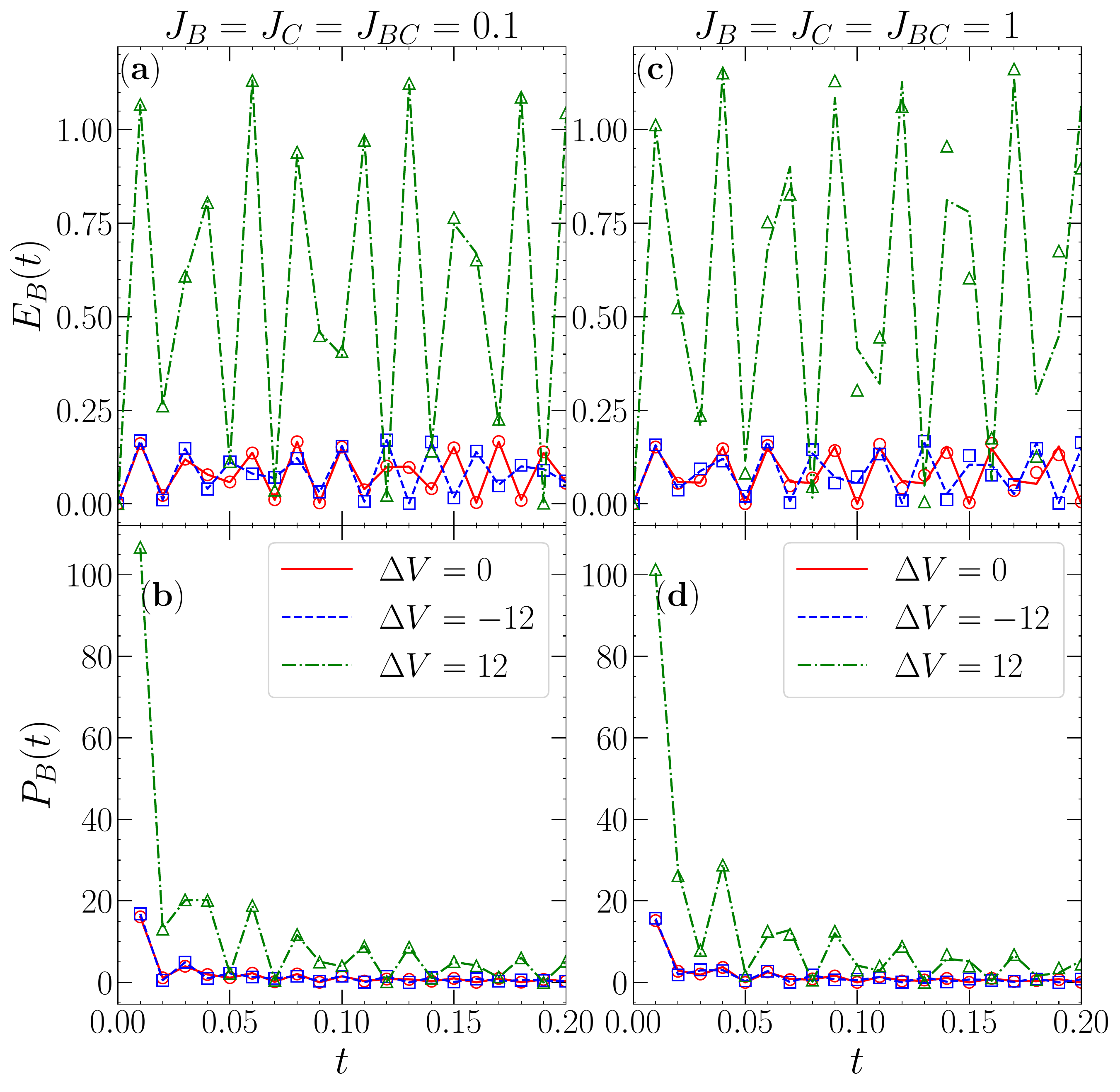}
    \end{center}
    \caption{Comparison of analytical and ED results for the dynamics of energy and power of the battery part; here $N_B=200, N_C=10,000$, with different Hamiltonian parameters as marked. Lines depict the exact numerical results whereas markers of the same color their corresponding analytical results.}
    \label{FIG7}
\end{figure}

As advanced in Sec.~\ref{subsec:large_spin}, the analytical results of the collective quantum battery under the Holstein-Primakoff transformation, valid at small excitation energies, are in good agreement with the ones from exact diagonalization, in particular at short-time scales. Figure.~\ref{FIG7} shows the comparison of the real-time dynamics of both the energy $E_B$ and power $P_B$ of the battery part, featuring $N_B=200$ and $N_C=10000$ qubits and various Hamiltonian parameters. The results exhibit a remarkable similarity, in particular at short time scales.

\begin{figure}[htbp]
    \begin{center}
    \includegraphics[width=0.99\columnwidth]{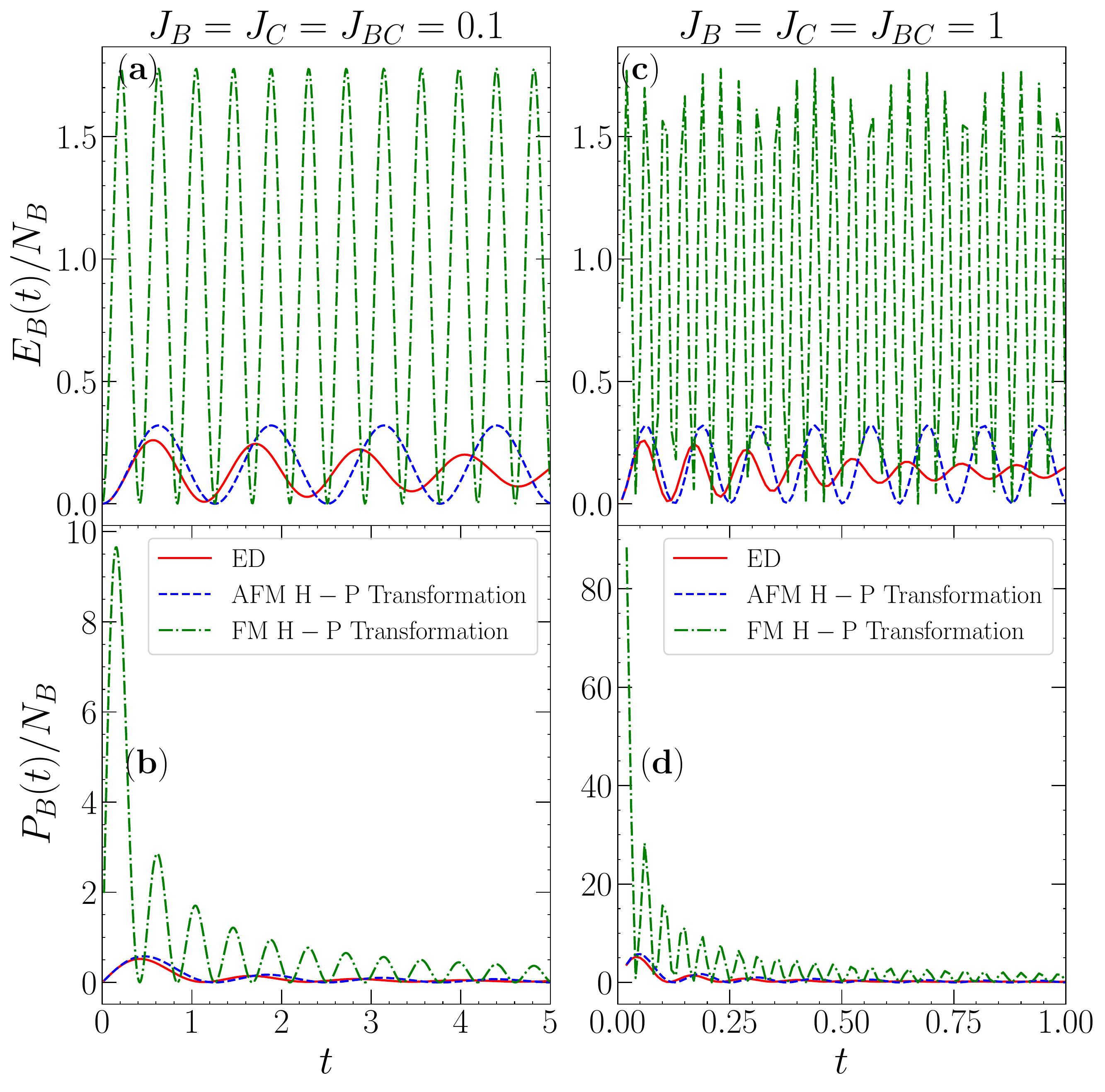}
    \end{center}
    \caption{Comparison of energy [(a) and (c)] and power [(b) and (d)] for either FM or AFM H-P transformations with $N_B=50, N_C=100, V_B=V_C=2$. Homogeneous couplings are chosen as $J=0.1$ and 1 for the left and right panels, respectively. Overall, the AFM H-P transformation gives better results when contrasted to the exact ones obtained via ED.}
    \label{FIG8}
\end{figure}

We promote a second benchmark, contrasting the results of both types of H-P transformation, FM and AFM, with the numerically exact ones in Fig.~\ref{FIG8}. Here it becomes clear that a transformation that follows the type of initial state we choose, $|\psi(0)\rangle = |\Downarrow \rangle_B\bigotimes |\Uparrow\rangle_C$, akin to an AFM large-spin state, fares better when compared to the ones stemming from ED. The FM H-P transformation results significantly miss not only the amplitude but also the frequency of dynamical oscillations, even at short-time scales.
\begin{figure}[htbp]
    \vskip 0.15in
    \begin{center}
    \includegraphics[width=0.75\columnwidth]{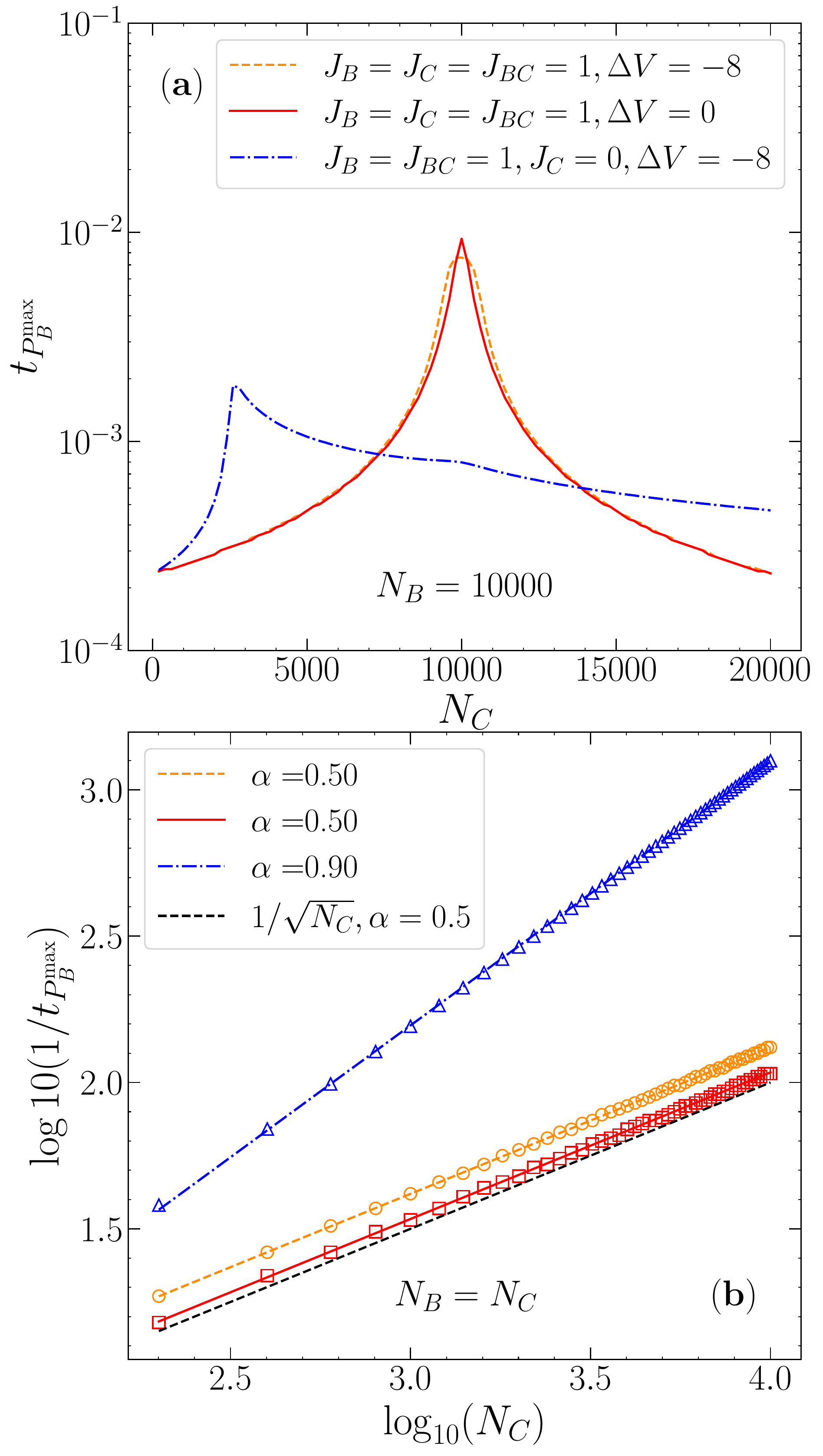}
    \end{center}
    \caption{Variations of $t_{P_B^{\rm max}}$, the corresponding time of $P_B^{\rm max}$. Two panels share the same configurations of Hamiltonian. (a) shows $t_{P_B^{\rm max}}$ of Fig.~\ref{FIG3}(d). One can see that, similar to other quantities, $t_{P_B^{\rm max}}$ of the homogeneous cases also shows peaks when $N_B\simeq N_C$, and for the inhomogeneous case such peak appears in small system size of the charger. (b) exhibits $t_{P_B^{\rm max}}$ of Fig.~\ref{FIG4}(d). Markers represent the numerical results and lines are the linear fitting to them. In contrast to the parallel case(black dashed line), $t_{P_B^{\rm max}}$ of the homogeneous cases also exhibits a scaling behavior of $1/\sqrt{N_C}$ while decreasing the strength of the couplings among the charger qubits makes it faster to reach the maxima of power.}
    \label{FIG9}
\end{figure}
\section{$t_{P_B^{\rm max}}$ of two settings \label{sec:t_Pmax}}
\indent We further check how $t_{P_B^{\rm max}}$ varies with the size of the charger for the two different settings investigated. Figure~\ref{FIG9}(a) shows the result of cases with $N_B$ fixed to 10,000, the same configurations of Hamiltonian as Fig.~\ref{FIG3} in the main text. Similar peaks are observed when $N_B\sim N_C$ for the homogeneous cases. Combined with the results of Fig.~\ref{FIG3}(a) and ~\ref{FIG3}(b) this states that an enhanced energy transfer is accompanied by a larger time it takes to accomplish it. In the inhomogeneous $J_C=0$ case, however, a peak appears when $N_C$ is much smaller than $N_B$, coinciding with the regime of a sharp increase of both $E_B^{\rm max}$ and $P_B^{\rm max}$ in Fig.~\ref{FIG3}.

Figure~\ref{FIG9}(b), on the other hand, shows the results of cases with $N_B=N_C$, the same configurations of Hamiltonian as Fig.~\ref{FIG4} in the main text; not the logarithmic scale in both axes. Taking the relation $1/t_{P_B^{\rm max}}\propto N_{\alpha}$, one can observe that the exponents of the homogeneous cases are the same as those of the special situation shown in Eq.~(\ref{eq:scaling}) of Sec.~\ref{subsec:large_spin}. For the inhomogeneous case, a diminished $J_C$ enhances the energy transfer from the charger to the battery part and makes the power reach its maximum faster with a larger scaling exponent.  
\section{Superextensive behaviors of noisy couplings \label{sec:noisy}}
\begin{figure}[htbp]
    \begin{center}
    \includegraphics[width=1.0\columnwidth]{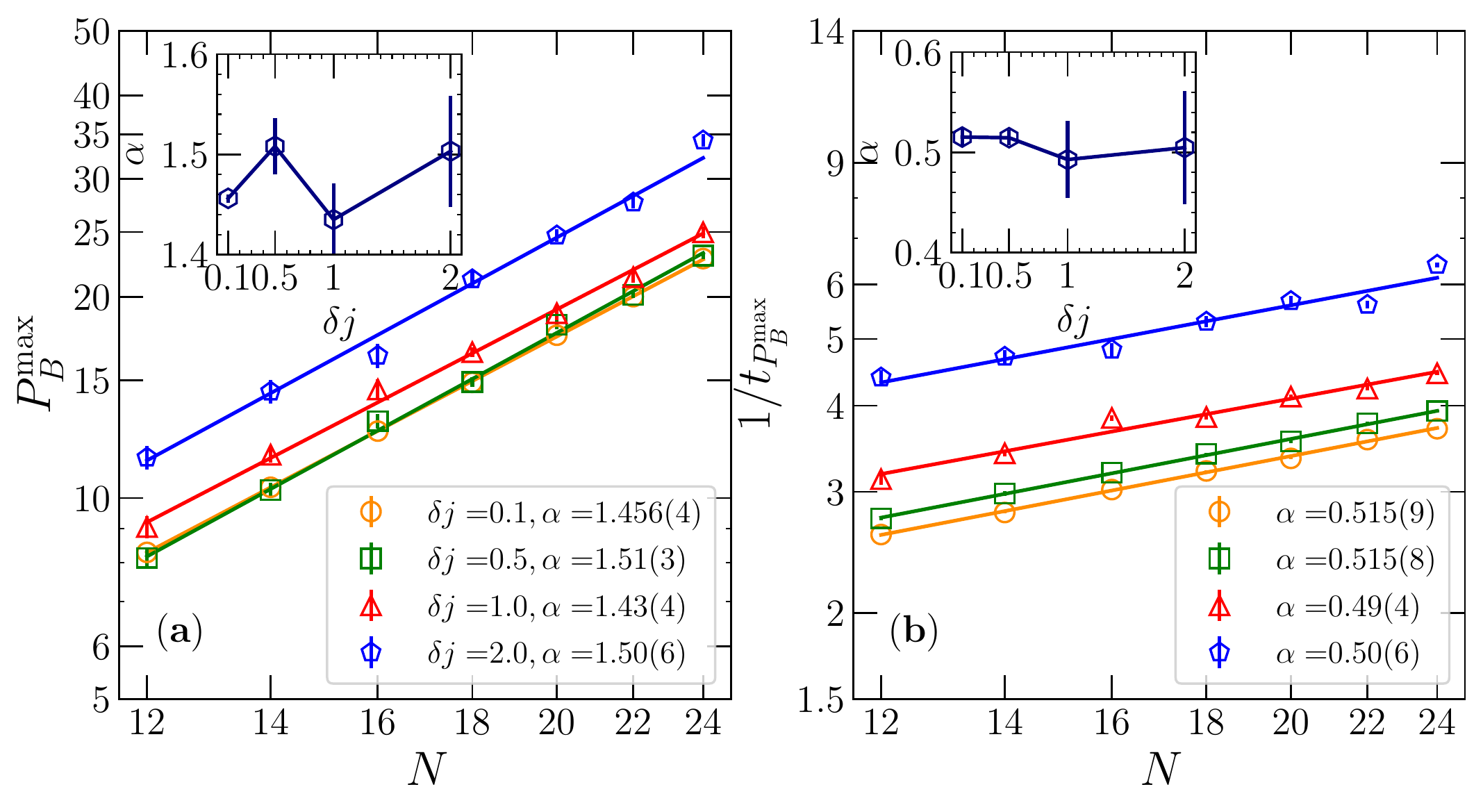}
    \end{center}
    \caption{\tcr{Collective battery with $J_{ij}$=1+$\delta$, where $\delta\in[-\delta j,\delta j]$ is the noise amplitude. Here, $\Delta V=0$ and 4 different magnitudes of $\delta j$ are investigated. There are 11 realizations for each $\delta j$; the error bar gives the standard error of the mean. The insets in both panels are how $\alpha$ changes with $\delta j$; variances of $\alpha$ serve as the error bar. These numerical results are derived by the Lanczos algorithm. Scaling exponents are close to those of the homogeneous cases in Fig.~\ref{FIG5}(d). One can see that cases with noise in couplings, which is closer to situations of experiments, still show robust superextensive behavior.}}
    \label{FIG10}
\end{figure}

\indent As a complement of Fig.~\ref{FIG4} in the main text, we further check by time-dependent Lanczos algorithm the cases when couplings acquire \tcr{different magnitude of noise} (details are given in the caption), which is closer to a more realistic situation of experiments. Figure~\ref{FIG10} shows the results: \tcr{Though} introducing noise to the couplings, the scaling exponents are remarkably close to those of the homogeneous cases with \tcr{$N_B=N_C=\frac{N}{2}$}, where a large spin representation is applicable and implies that the superextensive behavior observed in Fig.~\ref{FIG5} is robust to imperfections.


\bibliography{reference.bib}

\begin{thebibliography}{49}%
\makeatletter
\providecommand \@ifxundefined [1]{%
 \@ifx{#1\undefined}
}%
\providecommand \@ifnum [1]{%
 \ifnum #1\expandafter \@firstoftwo
 \else \expandafter \@secondoftwo
 \fi
}%
\providecommand \@ifx [1]{%
 \ifx #1\expandafter \@firstoftwo
 \else \expandafter \@secondoftwo
 \fi
}%
\providecommand \natexlab [1]{#1}%
\providecommand \enquote  [1]{``#1''}%
\providecommand \bibnamefont  [1]{#1}%
\providecommand \bibfnamefont [1]{#1}%
\providecommand \citenamefont [1]{#1}%
\providecommand \href@noop [0]{\@secondoftwo}%
\providecommand \href [0]{\begingroup \@sanitize@url \@href}%
\providecommand \@href[1]{\@@startlink{#1}\@@href}%
\providecommand \@@href[1]{\endgroup#1\@@endlink}%
\providecommand \@sanitize@url [0]{\catcode `\\12\catcode `\$12\catcode
  `\&12\catcode `\#12\catcode `\^12\catcode `\_12\catcode `\%12\relax}%
\providecommand \@@startlink[1]{}%
\providecommand \@@endlink[0]{}%
\providecommand \url  [0]{\begingroup\@sanitize@url \@url }%
\providecommand \@url [1]{\endgroup\@href {#1}{\urlprefix }}%
\providecommand \urlprefix  [0]{URL }%
\providecommand \Eprint [0]{\href }%
\providecommand \doibase [0]{http://dx.doi.org/}%
\providecommand \selectlanguage [0]{\@gobble}%
\providecommand \bibinfo  [0]{\@secondoftwo}%
\providecommand \bibfield  [0]{\@secondoftwo}%
\providecommand \translation [1]{[#1]}%
\providecommand \BibitemOpen [0]{}%
\providecommand \bibitemStop [0]{}%
\providecommand \bibitemNoStop [0]{.\EOS\space}%
\providecommand \EOS [0]{\spacefactor3000\relax}%
\providecommand \BibitemShut  [1]{\csname bibitem#1\endcsname}%
\let\auto@bib@innerbib\@empty
\bibitem [{\citenamefont {Duan}\ \emph {et~al.}(2001)\citenamefont {Duan},
  \citenamefont {Lukin}, \citenamefont {Cirac},\ and\ \citenamefont
  {Zoller}}]{ET_2001_NATURE_Zoller}%
  \BibitemOpen
  \bibfield  {author} {\bibinfo {author} {\bibfnamefont {L.~M.}\ \bibnamefont
  {Duan}}, \bibinfo {author} {\bibfnamefont {M.~D.}\ \bibnamefont {Lukin}},
  \bibinfo {author} {\bibfnamefont {J.~I.}\ \bibnamefont {Cirac}}, \ and\
  \bibinfo {author} {\bibfnamefont {P.}~\bibnamefont {Zoller}},\ }\href
  {\doibase 10.1038/35106500} {\bibfield  {journal} {\bibinfo  {journal}
  {Nature}\ }\textbf {\bibinfo {volume} {414}},\ \bibinfo {pages} {413}
  (\bibinfo {year} {2001})}\BibitemShut {NoStop}%
\bibitem [{\citenamefont {Fang}\ \emph {et~al.}(2020)\citenamefont {Fang},
  \citenamefont {Zeng}, \citenamefont {Liu}, \citenamefont {Zou}, \citenamefont
  {Wu}, \citenamefont {Tang}, \citenamefont {Sheng}, \citenamefont {Xiang},
  \citenamefont {Zhang}, \citenamefont {Li}, \citenamefont {Wang},
  \citenamefont {You}, \citenamefont {Li}, \citenamefont {Chen}, \citenamefont
  {Chen}, \citenamefont {Zhang}, \citenamefont {Peng}, \citenamefont {Ma},
  \citenamefont {Chen},\ and\ \citenamefont {Pan}}]{ET_2020_NatPho_Pan}%
  \BibitemOpen
  \bibfield  {author} {\bibinfo {author} {\bibfnamefont {X.-T.}\ \bibnamefont
  {Fang}}, \bibinfo {author} {\bibfnamefont {P.}~\bibnamefont {Zeng}}, \bibinfo
  {author} {\bibfnamefont {H.}~\bibnamefont {Liu}}, \bibinfo {author}
  {\bibfnamefont {M.}~\bibnamefont {Zou}}, \bibinfo {author} {\bibfnamefont
  {W.}~\bibnamefont {Wu}}, \bibinfo {author} {\bibfnamefont {Y.-L.}\
  \bibnamefont {Tang}}, \bibinfo {author} {\bibfnamefont {Y.-J.}\ \bibnamefont
  {Sheng}}, \bibinfo {author} {\bibfnamefont {Y.}~\bibnamefont {Xiang}},
  \bibinfo {author} {\bibfnamefont {W.}~\bibnamefont {Zhang}}, \bibinfo
  {author} {\bibfnamefont {H.}~\bibnamefont {Li}}, \bibinfo {author}
  {\bibfnamefont {Z.}~\bibnamefont {Wang}}, \bibinfo {author} {\bibfnamefont
  {L.}~\bibnamefont {You}}, \bibinfo {author} {\bibfnamefont {M.-J.}\
  \bibnamefont {Li}}, \bibinfo {author} {\bibfnamefont {H.}~\bibnamefont
  {Chen}}, \bibinfo {author} {\bibfnamefont {Y.-A.}\ \bibnamefont {Chen}},
  \bibinfo {author} {\bibfnamefont {Q.}~\bibnamefont {Zhang}}, \bibinfo
  {author} {\bibfnamefont {C.-Z.}\ \bibnamefont {Peng}}, \bibinfo {author}
  {\bibfnamefont {X.}~\bibnamefont {Ma}}, \bibinfo {author} {\bibfnamefont
  {T.-Y.}\ \bibnamefont {Chen}}, \ and\ \bibinfo {author} {\bibfnamefont
  {J.-W.}\ \bibnamefont {Pan}},\ }\href {\doibase 10.1038/s41566-020-0599-8}
  {\bibfield  {journal} {\bibinfo  {journal} {Nature Photonics}\ }\textbf
  {\bibinfo {volume} {14}},\ \bibinfo {pages} {422} (\bibinfo {year}
  {2020})}\BibitemShut {NoStop}%
\bibitem [{\citenamefont {Yin}\ \emph {et~al.}(2020)\citenamefont {Yin},
  \citenamefont {Li}, \citenamefont {Liao}, \citenamefont {Yang}, \citenamefont
  {Cao}, \citenamefont {Zhang}, \citenamefont {Ren}, \citenamefont {Cai},
  \citenamefont {Liu}, \citenamefont {Li}, \citenamefont {Shu}, \citenamefont
  {Huang}, \citenamefont {Deng}, \citenamefont {Li}, \citenamefont {Zhang},
  \citenamefont {Liu}, \citenamefont {Chen}, \citenamefont {Lu}, \citenamefont
  {Wang}, \citenamefont {Xu}, \citenamefont {Wang}, \citenamefont {Peng},
  \citenamefont {Ekert},\ and\ \citenamefont {Pan}}]{ET_2020_NATURE_Jian-Wei}%
  \BibitemOpen
  \bibfield  {author} {\bibinfo {author} {\bibfnamefont {J.}~\bibnamefont
  {Yin}}, \bibinfo {author} {\bibfnamefont {Y.-H.}\ \bibnamefont {Li}},
  \bibinfo {author} {\bibfnamefont {S.-K.}\ \bibnamefont {Liao}}, \bibinfo
  {author} {\bibfnamefont {M.}~\bibnamefont {Yang}}, \bibinfo {author}
  {\bibfnamefont {Y.}~\bibnamefont {Cao}}, \bibinfo {author} {\bibfnamefont
  {L.}~\bibnamefont {Zhang}}, \bibinfo {author} {\bibfnamefont {J.-G.}\
  \bibnamefont {Ren}}, \bibinfo {author} {\bibfnamefont {W.-Q.}\ \bibnamefont
  {Cai}}, \bibinfo {author} {\bibfnamefont {W.-Y.}\ \bibnamefont {Liu}},
  \bibinfo {author} {\bibfnamefont {S.-L.}\ \bibnamefont {Li}}, \bibinfo
  {author} {\bibfnamefont {R.}~\bibnamefont {Shu}}, \bibinfo {author}
  {\bibfnamefont {Y.-M.}\ \bibnamefont {Huang}}, \bibinfo {author}
  {\bibfnamefont {L.}~\bibnamefont {Deng}}, \bibinfo {author} {\bibfnamefont
  {L.}~\bibnamefont {Li}}, \bibinfo {author} {\bibfnamefont {Q.}~\bibnamefont
  {Zhang}}, \bibinfo {author} {\bibfnamefont {N.-L.}\ \bibnamefont {Liu}},
  \bibinfo {author} {\bibfnamefont {Y.-A.}\ \bibnamefont {Chen}}, \bibinfo
  {author} {\bibfnamefont {C.-Y.}\ \bibnamefont {Lu}}, \bibinfo {author}
  {\bibfnamefont {X.-B.}\ \bibnamefont {Wang}}, \bibinfo {author}
  {\bibfnamefont {F.}~\bibnamefont {Xu}}, \bibinfo {author} {\bibfnamefont
  {J.-Y.}\ \bibnamefont {Wang}}, \bibinfo {author} {\bibfnamefont {C.-Z.}\
  \bibnamefont {Peng}}, \bibinfo {author} {\bibfnamefont {A.~K.}\ \bibnamefont
  {Ekert}}, \ and\ \bibinfo {author} {\bibfnamefont {J.-W.}\ \bibnamefont
  {Pan}},\ }\href {\doibase 10.1038/s41586-020-2401-y} {\bibfield  {journal}
  {\bibinfo  {journal} {Nature}\ }\textbf {\bibinfo {volume} {582}},\ \bibinfo
  {pages} {501} (\bibinfo {year} {2020})}\BibitemShut {NoStop}%
\bibitem [{\citenamefont {Chen}\ \emph {et~al.}(2021)\citenamefont {Chen},
  \citenamefont {Zhang}, \citenamefont {Chen}, \citenamefont {Cai},
  \citenamefont {Liao}, \citenamefont {Zhang}, \citenamefont {Chen},
  \citenamefont {Yin}, \citenamefont {Ren}, \citenamefont {Chen}, \citenamefont
  {Han}, \citenamefont {Yu}, \citenamefont {Liang}, \citenamefont {Zhou},
  \citenamefont {Yuan}, \citenamefont {Zhao}, \citenamefont {Wang},
  \citenamefont {Jiang}, \citenamefont {Zhang}, \citenamefont {Liu},
  \citenamefont {Li}, \citenamefont {Shen}, \citenamefont {Cao}, \citenamefont
  {Lu}, \citenamefont {Shu}, \citenamefont {Wang}, \citenamefont {Li},
  \citenamefont {Liu}, \citenamefont {Xu}, \citenamefont {Wang}, \citenamefont
  {Peng},\ and\ \citenamefont {Pan}}]{ET_2021_NATURE_Pan}%
  \BibitemOpen
  \bibfield  {author} {\bibinfo {author} {\bibfnamefont {Y.-A.}\ \bibnamefont
  {Chen}}, \bibinfo {author} {\bibfnamefont {Q.}~\bibnamefont {Zhang}},
  \bibinfo {author} {\bibfnamefont {T.-Y.}\ \bibnamefont {Chen}}, \bibinfo
  {author} {\bibfnamefont {W.-Q.}\ \bibnamefont {Cai}}, \bibinfo {author}
  {\bibfnamefont {S.-K.}\ \bibnamefont {Liao}}, \bibinfo {author}
  {\bibfnamefont {J.}~\bibnamefont {Zhang}}, \bibinfo {author} {\bibfnamefont
  {K.}~\bibnamefont {Chen}}, \bibinfo {author} {\bibfnamefont {J.}~\bibnamefont
  {Yin}}, \bibinfo {author} {\bibfnamefont {J.-G.}\ \bibnamefont {Ren}},
  \bibinfo {author} {\bibfnamefont {Z.}~\bibnamefont {Chen}}, \bibinfo {author}
  {\bibfnamefont {S.-L.}\ \bibnamefont {Han}}, \bibinfo {author} {\bibfnamefont
  {Q.}~\bibnamefont {Yu}}, \bibinfo {author} {\bibfnamefont {K.}~\bibnamefont
  {Liang}}, \bibinfo {author} {\bibfnamefont {F.}~\bibnamefont {Zhou}},
  \bibinfo {author} {\bibfnamefont {X.}~\bibnamefont {Yuan}}, \bibinfo {author}
  {\bibfnamefont {M.-S.}\ \bibnamefont {Zhao}}, \bibinfo {author}
  {\bibfnamefont {T.-Y.}\ \bibnamefont {Wang}}, \bibinfo {author}
  {\bibfnamefont {X.}~\bibnamefont {Jiang}}, \bibinfo {author} {\bibfnamefont
  {L.}~\bibnamefont {Zhang}}, \bibinfo {author} {\bibfnamefont {W.-Y.}\
  \bibnamefont {Liu}}, \bibinfo {author} {\bibfnamefont {Y.}~\bibnamefont
  {Li}}, \bibinfo {author} {\bibfnamefont {Q.}~\bibnamefont {Shen}}, \bibinfo
  {author} {\bibfnamefont {Y.}~\bibnamefont {Cao}}, \bibinfo {author}
  {\bibfnamefont {C.-Y.}\ \bibnamefont {Lu}}, \bibinfo {author} {\bibfnamefont
  {R.}~\bibnamefont {Shu}}, \bibinfo {author} {\bibfnamefont {J.-Y.}\
  \bibnamefont {Wang}}, \bibinfo {author} {\bibfnamefont {L.}~\bibnamefont
  {Li}}, \bibinfo {author} {\bibfnamefont {N.-L.}\ \bibnamefont {Liu}},
  \bibinfo {author} {\bibfnamefont {F.}~\bibnamefont {Xu}}, \bibinfo {author}
  {\bibfnamefont {X.-B.}\ \bibnamefont {Wang}}, \bibinfo {author}
  {\bibfnamefont {C.-Z.}\ \bibnamefont {Peng}}, \ and\ \bibinfo {author}
  {\bibfnamefont {J.-W.}\ \bibnamefont {Pan}},\ }\href {\doibase
  10.1038/s41586-020-03093-8} {\bibfield  {journal} {\bibinfo  {journal}
  {Nature}\ }\textbf {\bibinfo {volume} {589}},\ \bibinfo {pages} {214}
  (\bibinfo {year} {2021})}\BibitemShut {NoStop}%
\bibitem [{\citenamefont {Ladd}\ \emph {et~al.}(2010)\citenamefont {Ladd},
  \citenamefont {Jelezko}, \citenamefont {Laflamme}, \citenamefont {Nakamura},
  \citenamefont {Monroe},\ and\ \citenamefont
  {O’Brien}}]{ET_2010_NATURE_Brien}%
  \BibitemOpen
  \bibfield  {author} {\bibinfo {author} {\bibfnamefont {T.~D.}\ \bibnamefont
  {Ladd}}, \bibinfo {author} {\bibfnamefont {F.}~\bibnamefont {Jelezko}},
  \bibinfo {author} {\bibfnamefont {R.}~\bibnamefont {Laflamme}}, \bibinfo
  {author} {\bibfnamefont {Y.}~\bibnamefont {Nakamura}}, \bibinfo {author}
  {\bibfnamefont {C.}~\bibnamefont {Monroe}}, \ and\ \bibinfo {author}
  {\bibfnamefont {J.~L.}\ \bibnamefont {O’Brien}},\ }\href {\doibase
  10.1038/nature08812} {\bibfield  {journal} {\bibinfo  {journal} {Nature}\
  }\textbf {\bibinfo {volume} {464}},\ \bibinfo {pages} {45} (\bibinfo {year}
  {2010})}\BibitemShut {NoStop}%
\bibitem [{\citenamefont {Reiher}\ \emph {et~al.}(2017)\citenamefont {Reiher},
  \citenamefont {Wiebe}, \citenamefont {Svore}, \citenamefont {Wecker},\ and\
  \citenamefont {Troyer}}]{T_2017_PNAS_Matthias}%
  \BibitemOpen
  \bibfield  {author} {\bibinfo {author} {\bibfnamefont {M.}~\bibnamefont
  {Reiher}}, \bibinfo {author} {\bibfnamefont {N.}~\bibnamefont {Wiebe}},
  \bibinfo {author} {\bibfnamefont {K.~M.}\ \bibnamefont {Svore}}, \bibinfo
  {author} {\bibfnamefont {D.}~\bibnamefont {Wecker}}, \ and\ \bibinfo {author}
  {\bibfnamefont {M.}~\bibnamefont {Troyer}},\ }\href {\doibase
  10.1073/pnas.1619152114} {\bibfield  {journal} {\bibinfo  {journal}
  {Proceedings of the National Academy of Sciences}\ }\textbf {\bibinfo
  {volume} {114}},\ \bibinfo {pages} {7555} (\bibinfo {year}
  {2017})}\BibitemShut {NoStop}%
\bibitem [{\citenamefont {Proctor}\ \emph {et~al.}(2021)\citenamefont
  {Proctor}, \citenamefont {Rudinger}, \citenamefont {Young}, \citenamefont
  {Nielsen},\ and\ \citenamefont {Blume-Kohout}}]{ET_2021_NatPhy_Robin}%
  \BibitemOpen
  \bibfield  {author} {\bibinfo {author} {\bibfnamefont {T.}~\bibnamefont
  {Proctor}}, \bibinfo {author} {\bibfnamefont {K.}~\bibnamefont {Rudinger}},
  \bibinfo {author} {\bibfnamefont {K.}~\bibnamefont {Young}}, \bibinfo
  {author} {\bibfnamefont {E.}~\bibnamefont {Nielsen}}, \ and\ \bibinfo
  {author} {\bibfnamefont {R.}~\bibnamefont {Blume-Kohout}},\ }\href {\doibase
  10.1038/s41567-021-01409-7} {\bibfield  {journal} {\bibinfo  {journal}
  {Nature Physics}\ } (\bibinfo {year} {2021}),\
  10.1038/s41567-021-01409-7}\BibitemShut {NoStop}%
\bibitem [{\citenamefont {Koepsell}\ \emph {et~al.}(2021)\citenamefont
  {Koepsell}, \citenamefont {Bourgund}, \citenamefont {Sompet}, \citenamefont
  {Hirthe}, \citenamefont {Bohrdt}, \citenamefont {Wang}, \citenamefont
  {Grusdt}, \citenamefont {Demler}, \citenamefont {Salomon}, \citenamefont
  {Gross},\ and\ \citenamefont {Bloch}}]{TE_2021_SCIENCE_Bloch}%
  \BibitemOpen
  \bibfield  {author} {\bibinfo {author} {\bibfnamefont {J.}~\bibnamefont
  {Koepsell}}, \bibinfo {author} {\bibfnamefont {D.}~\bibnamefont {Bourgund}},
  \bibinfo {author} {\bibfnamefont {P.}~\bibnamefont {Sompet}}, \bibinfo
  {author} {\bibfnamefont {S.}~\bibnamefont {Hirthe}}, \bibinfo {author}
  {\bibfnamefont {A.}~\bibnamefont {Bohrdt}}, \bibinfo {author} {\bibfnamefont
  {Y.}~\bibnamefont {Wang}}, \bibinfo {author} {\bibfnamefont {F.}~\bibnamefont
  {Grusdt}}, \bibinfo {author} {\bibfnamefont {E.}~\bibnamefont {Demler}},
  \bibinfo {author} {\bibfnamefont {G.}~\bibnamefont {Salomon}}, \bibinfo
  {author} {\bibfnamefont {C.}~\bibnamefont {Gross}}, \ and\ \bibinfo {author}
  {\bibfnamefont {I.}~\bibnamefont {Bloch}},\ }\href {\doibase
  10.1126/science.abe7165} {\bibfield  {journal} {\bibinfo  {journal}
  {Science}\ }\textbf {\bibinfo {volume} {374}},\ \bibinfo {pages} {82}
  (\bibinfo {year} {2021})}\BibitemShut {NoStop}%
\bibitem [{\citenamefont {Brown}\ \emph {et~al.}(2019)\citenamefont {Brown},
  \citenamefont {Mitra}, \citenamefont {Guardado-Sanchez}, \citenamefont
  {Nourafkan}, \citenamefont {Reymbaut}, \citenamefont {Hébert}, \citenamefont
  {Bergeron}, \citenamefont {Tremblay}, \citenamefont {Kokalj}, \citenamefont
  {Huse}, \citenamefont {Schauß},\ and\ \citenamefont
  {Bakr}}]{ET_2019_SCIENCE_Bark}%
  \BibitemOpen
  \bibfield  {author} {\bibinfo {author} {\bibfnamefont {P.~T.}\ \bibnamefont
  {Brown}}, \bibinfo {author} {\bibfnamefont {D.}~\bibnamefont {Mitra}},
  \bibinfo {author} {\bibfnamefont {E.}~\bibnamefont {Guardado-Sanchez}},
  \bibinfo {author} {\bibfnamefont {R.}~\bibnamefont {Nourafkan}}, \bibinfo
  {author} {\bibfnamefont {A.}~\bibnamefont {Reymbaut}}, \bibinfo {author}
  {\bibfnamefont {C.-D.}\ \bibnamefont {Hébert}}, \bibinfo {author}
  {\bibfnamefont {S.}~\bibnamefont {Bergeron}}, \bibinfo {author}
  {\bibfnamefont {A.-M.~S.}\ \bibnamefont {Tremblay}}, \bibinfo {author}
  {\bibfnamefont {J.}~\bibnamefont {Kokalj}}, \bibinfo {author} {\bibfnamefont
  {D.~A.}\ \bibnamefont {Huse}}, \bibinfo {author} {\bibfnamefont
  {P.}~\bibnamefont {Schauß}}, \ and\ \bibinfo {author} {\bibfnamefont
  {W.~S.}\ \bibnamefont {Bakr}},\ }\href {\doibase 10.1126/science.aat4134}
  {\bibfield  {journal} {\bibinfo  {journal} {Science}\ }\textbf {\bibinfo
  {volume} {363}},\ \bibinfo {pages} {379} (\bibinfo {year} {2019})},\ \Eprint
  {http://arxiv.org/abs/https://www.science.org/doi/pdf/10.1126/science.aat4134}
  {https://www.science.org/doi/pdf/10.1126/science.aat4134} \BibitemShut
  {NoStop}%
\bibitem [{\citenamefont {Mil}\ \emph {et~al.}(2020)\citenamefont {Mil},
  \citenamefont {Zache}, \citenamefont {Hegde}, \citenamefont {Xia},
  \citenamefont {Bhatt}, \citenamefont {Oberthaler}, \citenamefont {Hauke},
  \citenamefont {Berges},\ and\ \citenamefont
  {Jendrzejewski}}]{ET_2020_SCIENCE_Jendrzejewski}%
  \BibitemOpen
  \bibfield  {author} {\bibinfo {author} {\bibfnamefont {A.}~\bibnamefont
  {Mil}}, \bibinfo {author} {\bibfnamefont {T.~V.}\ \bibnamefont {Zache}},
  \bibinfo {author} {\bibfnamefont {A.}~\bibnamefont {Hegde}}, \bibinfo
  {author} {\bibfnamefont {A.}~\bibnamefont {Xia}}, \bibinfo {author}
  {\bibfnamefont {R.~P.}\ \bibnamefont {Bhatt}}, \bibinfo {author}
  {\bibfnamefont {M.~K.}\ \bibnamefont {Oberthaler}}, \bibinfo {author}
  {\bibfnamefont {P.}~\bibnamefont {Hauke}}, \bibinfo {author} {\bibfnamefont
  {J.}~\bibnamefont {Berges}}, \ and\ \bibinfo {author} {\bibfnamefont
  {F.}~\bibnamefont {Jendrzejewski}},\ }\href {\doibase
  10.1126/science.aaz5312} {\bibfield  {journal} {\bibinfo  {journal}
  {Science}\ }\textbf {\bibinfo {volume} {367}},\ \bibinfo {pages} {1128}
  (\bibinfo {year} {2020})}\BibitemShut {NoStop}%
\bibitem [{\citenamefont {Muniz}\ \emph {et~al.}(2020)\citenamefont {Muniz},
  \citenamefont {Barberena}, \citenamefont {Lewis-Swan}, \citenamefont {Young},
  \citenamefont {Cline}, \citenamefont {Rey},\ and\ \citenamefont
  {Thompson}}]{ET_2020_NATURE_Thompson}%
  \BibitemOpen
  \bibfield  {author} {\bibinfo {author} {\bibfnamefont {J.~A.}\ \bibnamefont
  {Muniz}}, \bibinfo {author} {\bibfnamefont {D.}~\bibnamefont {Barberena}},
  \bibinfo {author} {\bibfnamefont {R.~J.}\ \bibnamefont {Lewis-Swan}},
  \bibinfo {author} {\bibfnamefont {D.~J.}\ \bibnamefont {Young}}, \bibinfo
  {author} {\bibfnamefont {J.~R.~K.}\ \bibnamefont {Cline}}, \bibinfo {author}
  {\bibfnamefont {A.~M.}\ \bibnamefont {Rey}}, \ and\ \bibinfo {author}
  {\bibfnamefont {J.~K.}\ \bibnamefont {Thompson}},\ }\href {\doibase
  10.1038/s41586-020-2224-x} {\bibfield  {journal} {\bibinfo  {journal}
  {Nature}\ }\textbf {\bibinfo {volume} {580}},\ \bibinfo {pages} {602}
  (\bibinfo {year} {2020})}\BibitemShut {NoStop}%
\bibitem [{\citenamefont {Ebadi}\ \emph {et~al.}(2021)\citenamefont {Ebadi},
  \citenamefont {Wang}, \citenamefont {Levine}, \citenamefont {Keesling},
  \citenamefont {Semeghini}, \citenamefont {Omran}, \citenamefont {Bluvstein},
  \citenamefont {Samajdar}, \citenamefont {Pichler}, \citenamefont {Ho},
  \citenamefont {Choi}, \citenamefont {Sachdev}, \citenamefont {Greiner},
  \citenamefont {Vuletić},\ and\ \citenamefont
  {Lukin}}]{ET_2021_NATURE_Lukin}%
  \BibitemOpen
  \bibfield  {author} {\bibinfo {author} {\bibfnamefont {S.}~\bibnamefont
  {Ebadi}}, \bibinfo {author} {\bibfnamefont {T.~T.}\ \bibnamefont {Wang}},
  \bibinfo {author} {\bibfnamefont {H.}~\bibnamefont {Levine}}, \bibinfo
  {author} {\bibfnamefont {A.}~\bibnamefont {Keesling}}, \bibinfo {author}
  {\bibfnamefont {G.}~\bibnamefont {Semeghini}}, \bibinfo {author}
  {\bibfnamefont {A.}~\bibnamefont {Omran}}, \bibinfo {author} {\bibfnamefont
  {D.}~\bibnamefont {Bluvstein}}, \bibinfo {author} {\bibfnamefont
  {R.}~\bibnamefont {Samajdar}}, \bibinfo {author} {\bibfnamefont
  {H.}~\bibnamefont {Pichler}}, \bibinfo {author} {\bibfnamefont {W.~W.}\
  \bibnamefont {Ho}}, \bibinfo {author} {\bibfnamefont {S.}~\bibnamefont
  {Choi}}, \bibinfo {author} {\bibfnamefont {S.}~\bibnamefont {Sachdev}},
  \bibinfo {author} {\bibfnamefont {M.}~\bibnamefont {Greiner}}, \bibinfo
  {author} {\bibfnamefont {V.}~\bibnamefont {Vuletić}}, \ and\ \bibinfo
  {author} {\bibfnamefont {M.~D.}\ \bibnamefont {Lukin}},\ }\href {\doibase
  10.1038/s41586-021-03582-4} {\bibfield  {journal} {\bibinfo  {journal}
  {Nature}\ }\textbf {\bibinfo {volume} {595}},\ \bibinfo {pages} {227}
  (\bibinfo {year} {2021})}\BibitemShut {NoStop}%
\bibitem [{\citenamefont {Lanyon}\ \emph {et~al.}(2011)\citenamefont {Lanyon},
  \citenamefont {Hempel}, \citenamefont {Nigg}, \citenamefont {Müller},
  \citenamefont {Gerritsma}, \citenamefont {Zähringer}, \citenamefont
  {Schindler}, \citenamefont {Barreiro}, \citenamefont {Rambach}, \citenamefont
  {Kirchmair}, \citenamefont {Hennrich}, \citenamefont {Zoller}, \citenamefont
  {Blatt},\ and\ \citenamefont {Roos}}]{E_2011_SCIENCE_Lanyon}%
  \BibitemOpen
  \bibfield  {author} {\bibinfo {author} {\bibfnamefont {B.~P.}\ \bibnamefont
  {Lanyon}}, \bibinfo {author} {\bibfnamefont {C.}~\bibnamefont {Hempel}},
  \bibinfo {author} {\bibfnamefont {D.}~\bibnamefont {Nigg}}, \bibinfo {author}
  {\bibfnamefont {M.}~\bibnamefont {Müller}}, \bibinfo {author} {\bibfnamefont
  {R.}~\bibnamefont {Gerritsma}}, \bibinfo {author} {\bibfnamefont
  {F.}~\bibnamefont {Zähringer}}, \bibinfo {author} {\bibfnamefont
  {P.}~\bibnamefont {Schindler}}, \bibinfo {author} {\bibfnamefont {J.~T.}\
  \bibnamefont {Barreiro}}, \bibinfo {author} {\bibfnamefont {M.}~\bibnamefont
  {Rambach}}, \bibinfo {author} {\bibfnamefont {G.}~\bibnamefont {Kirchmair}},
  \bibinfo {author} {\bibfnamefont {M.}~\bibnamefont {Hennrich}}, \bibinfo
  {author} {\bibfnamefont {P.}~\bibnamefont {Zoller}}, \bibinfo {author}
  {\bibfnamefont {R.}~\bibnamefont {Blatt}}, \ and\ \bibinfo {author}
  {\bibfnamefont {C.~F.}\ \bibnamefont {Roos}},\ }\href {\doibase
  10.1126/science.1208001} {\bibfield  {journal} {\bibinfo  {journal}
  {Science}\ }\textbf {\bibinfo {volume} {334}},\ \bibinfo {pages} {57}
  (\bibinfo {year} {2011})},\ \Eprint
  {http://arxiv.org/abs/https://www.science.org/doi/pdf/10.1126/science.1208001}
  {https://www.science.org/doi/pdf/10.1126/science.1208001} \BibitemShut
  {NoStop}%
\bibitem [{\citenamefont {Blatt}\ and\ \citenamefont
  {Roos}(2012)}]{E_2012_NATURE_Blatt}%
  \BibitemOpen
  \bibfield  {author} {\bibinfo {author} {\bibfnamefont {R.}~\bibnamefont
  {Blatt}}\ and\ \bibinfo {author} {\bibfnamefont {C.~F.}\ \bibnamefont
  {Roos}},\ }\href {\doibase 10.1038/nphys2252} {\bibfield  {journal} {\bibinfo
   {journal} {Nature Physics}\ }\textbf {\bibinfo {volume} {8}},\ \bibinfo
  {pages} {277} (\bibinfo {year} {2012})}\BibitemShut {NoStop}%
\bibitem [{\citenamefont {Wolf}\ \emph {et~al.}(2019)\citenamefont {Wolf},
  \citenamefont {Shi}, \citenamefont {Heip}, \citenamefont {Gessner},
  \citenamefont {Pezzè}, \citenamefont {Smerzi}, \citenamefont {Schulte},
  \citenamefont {Hammerer},\ and\ \citenamefont
  {Schmidt}}]{E_2019_NatComm_Fabian}%
  \BibitemOpen
  \bibfield  {author} {\bibinfo {author} {\bibfnamefont {F.}~\bibnamefont
  {Wolf}}, \bibinfo {author} {\bibfnamefont {C.}~\bibnamefont {Shi}}, \bibinfo
  {author} {\bibfnamefont {J.~C.}\ \bibnamefont {Heip}}, \bibinfo {author}
  {\bibfnamefont {M.}~\bibnamefont {Gessner}}, \bibinfo {author} {\bibfnamefont
  {L.}~\bibnamefont {Pezzè}}, \bibinfo {author} {\bibfnamefont
  {A.}~\bibnamefont {Smerzi}}, \bibinfo {author} {\bibfnamefont
  {M.}~\bibnamefont {Schulte}}, \bibinfo {author} {\bibfnamefont
  {K.}~\bibnamefont {Hammerer}}, \ and\ \bibinfo {author} {\bibfnamefont
  {P.~O.}\ \bibnamefont {Schmidt}},\ }\href {\doibase
  10.1038/s41467-019-10576-4} {\bibfield  {journal} {\bibinfo  {journal}
  {Nature Communications}\ }\textbf {\bibinfo {volume} {10}},\ \bibinfo {pages}
  {2929} (\bibinfo {year} {2019})}\BibitemShut {NoStop}%
\bibitem [{\citenamefont {Tamura}\ \emph {et~al.}(2020)\citenamefont {Tamura},
  \citenamefont {Mukaiyama},\ and\ \citenamefont {Toyoda}}]{E_2020_PRL_Tamura}%
  \BibitemOpen
  \bibfield  {author} {\bibinfo {author} {\bibfnamefont {M.}~\bibnamefont
  {Tamura}}, \bibinfo {author} {\bibfnamefont {T.}~\bibnamefont {Mukaiyama}}, \
  and\ \bibinfo {author} {\bibfnamefont {K.}~\bibnamefont {Toyoda}},\ }\href
  {\doibase 10.1103/PhysRevLett.124.200501} {\bibfield  {journal} {\bibinfo
  {journal} {Phys. Rev. Lett.}\ }\textbf {\bibinfo {volume} {124}},\ \bibinfo
  {pages} {200501} (\bibinfo {year} {2020})}\BibitemShut {NoStop}%
\bibitem [{\citenamefont {Monroe}\ \emph {et~al.}(2021)\citenamefont {Monroe},
  \citenamefont {Campbell}, \citenamefont {Duan}, \citenamefont {Gong},
  \citenamefont {Gorshkov}, \citenamefont {Hess}, \citenamefont {Islam},
  \citenamefont {Kim}, \citenamefont {Linke}, \citenamefont {Pagano},
  \citenamefont {Richerme}, \citenamefont {Senko},\ and\ \citenamefont
  {Yao}}]{R_2021_RMP_Monroe}%
  \BibitemOpen
  \bibfield  {author} {\bibinfo {author} {\bibfnamefont {C.}~\bibnamefont
  {Monroe}}, \bibinfo {author} {\bibfnamefont {W.~C.}\ \bibnamefont
  {Campbell}}, \bibinfo {author} {\bibfnamefont {L.-M.}\ \bibnamefont {Duan}},
  \bibinfo {author} {\bibfnamefont {Z.-X.}\ \bibnamefont {Gong}}, \bibinfo
  {author} {\bibfnamefont {A.~V.}\ \bibnamefont {Gorshkov}}, \bibinfo {author}
  {\bibfnamefont {P.~W.}\ \bibnamefont {Hess}}, \bibinfo {author}
  {\bibfnamefont {R.}~\bibnamefont {Islam}}, \bibinfo {author} {\bibfnamefont
  {K.}~\bibnamefont {Kim}}, \bibinfo {author} {\bibfnamefont {N.~M.}\
  \bibnamefont {Linke}}, \bibinfo {author} {\bibfnamefont {G.}~\bibnamefont
  {Pagano}}, \bibinfo {author} {\bibfnamefont {P.}~\bibnamefont {Richerme}},
  \bibinfo {author} {\bibfnamefont {C.}~\bibnamefont {Senko}}, \ and\ \bibinfo
  {author} {\bibfnamefont {N.~Y.}\ \bibnamefont {Yao}},\ }\href {\doibase
  10.1103/RevModPhys.93.025001} {\bibfield  {journal} {\bibinfo  {journal}
  {Rev. Mod. Phys.}\ }\textbf {\bibinfo {volume} {93}},\ \bibinfo {pages}
  {025001} (\bibinfo {year} {2021})}\BibitemShut {NoStop}%
\bibitem [{\citenamefont {Guo}\ \emph {et~al.}(2021)\citenamefont {Guo},
  \citenamefont {Cheng}, \citenamefont {Sun}, \citenamefont {Song},
  \citenamefont {Li}, \citenamefont {Wang}, \citenamefont {Ren}, \citenamefont
  {Dong}, \citenamefont {Zheng}, \citenamefont {Zhang}, \citenamefont
  {Mondaini}, \citenamefont {Fan},\ and\ \citenamefont
  {Wang}}]{TE_2021_NatPhy}%
  \BibitemOpen
  \bibfield  {author} {\bibinfo {author} {\bibfnamefont {Q.}~\bibnamefont
  {Guo}}, \bibinfo {author} {\bibfnamefont {C.}~\bibnamefont {Cheng}}, \bibinfo
  {author} {\bibfnamefont {Z.-H.}\ \bibnamefont {Sun}}, \bibinfo {author}
  {\bibfnamefont {Z.}~\bibnamefont {Song}}, \bibinfo {author} {\bibfnamefont
  {H.}~\bibnamefont {Li}}, \bibinfo {author} {\bibfnamefont {Z.}~\bibnamefont
  {Wang}}, \bibinfo {author} {\bibfnamefont {W.}~\bibnamefont {Ren}}, \bibinfo
  {author} {\bibfnamefont {H.}~\bibnamefont {Dong}}, \bibinfo {author}
  {\bibfnamefont {D.}~\bibnamefont {Zheng}}, \bibinfo {author} {\bibfnamefont
  {Y.-R.}\ \bibnamefont {Zhang}}, \bibinfo {author} {\bibfnamefont
  {R.}~\bibnamefont {Mondaini}}, \bibinfo {author} {\bibfnamefont
  {H.}~\bibnamefont {Fan}}, \ and\ \bibinfo {author} {\bibfnamefont
  {H.}~\bibnamefont {Wang}},\ }\href {\doibase 10.1038/s41567-020-1035-1}
  {\bibfield  {journal} {\bibinfo  {journal} {Nature Physics}\ }\textbf
  {\bibinfo {volume} {17}},\ \bibinfo {pages} {234} (\bibinfo {year}
  {2021})}\BibitemShut {NoStop}%
\bibitem [{\citenamefont {Niu}\ \emph {et~al.}(2021)\citenamefont {Niu},
  \citenamefont {Yan}, \citenamefont {Zhou}, \citenamefont {Tao}, \citenamefont
  {Li}, \citenamefont {Liu}, \citenamefont {Zhang}, \citenamefont {Jia},
  \citenamefont {Liu}, \citenamefont {Yan}, \citenamefont {Chen},\ and\
  \citenamefont {Yu}}]{ET_2021_ScienceBulletin_Yu}%
  \BibitemOpen
  \bibfield  {author} {\bibinfo {author} {\bibfnamefont {J.}~\bibnamefont
  {Niu}}, \bibinfo {author} {\bibfnamefont {T.}~\bibnamefont {Yan}}, \bibinfo
  {author} {\bibfnamefont {Y.}~\bibnamefont {Zhou}}, \bibinfo {author}
  {\bibfnamefont {Z.}~\bibnamefont {Tao}}, \bibinfo {author} {\bibfnamefont
  {X.}~\bibnamefont {Li}}, \bibinfo {author} {\bibfnamefont {W.}~\bibnamefont
  {Liu}}, \bibinfo {author} {\bibfnamefont {L.}~\bibnamefont {Zhang}}, \bibinfo
  {author} {\bibfnamefont {H.}~\bibnamefont {Jia}}, \bibinfo {author}
  {\bibfnamefont {S.}~\bibnamefont {Liu}}, \bibinfo {author} {\bibfnamefont
  {Z.}~\bibnamefont {Yan}}, \bibinfo {author} {\bibfnamefont {Y.}~\bibnamefont
  {Chen}}, \ and\ \bibinfo {author} {\bibfnamefont {D.}~\bibnamefont {Yu}},\
  }\href {\doibase https://doi.org/10.1016/j.scib.2021.02.035} {\bibfield
  {journal} {\bibinfo  {journal} {Science Bulletin}\ }\textbf {\bibinfo
  {volume} {66}},\ \bibinfo {pages} {1168} (\bibinfo {year}
  {2021})}\BibitemShut {NoStop}%
\bibitem [{\citenamefont {QUANTUM}\ \emph {et~al.}(2020)\citenamefont
  {QUANTUM}, \citenamefont {COLLABORATORS}, \citenamefont {Arute},
  \citenamefont {Arya}, \citenamefont {Babbush}, \citenamefont {Bacon},
  \citenamefont {Bardin}, \citenamefont {Barends}, \citenamefont {Boixo},
  \citenamefont {Broughton}, \citenamefont {Buckley}, \citenamefont {Buell},
  \citenamefont {Burkett}, \citenamefont {Bushnell}, \citenamefont {Chen},
  \citenamefont {Chen}, \citenamefont {Chiaro}, \citenamefont {Collins},
  \citenamefont {Courtney}, \citenamefont {Demura}, \citenamefont {Dunsworth},
  \citenamefont {Farhi}, \citenamefont {Fowler}, \citenamefont {Foxen},
  \citenamefont {Gidney}, \citenamefont {Giustina}, \citenamefont {Graff},
  \citenamefont {Habegger}, \citenamefont {Harrigan}, \citenamefont {Ho},
  \citenamefont {Hong}, \citenamefont {Huang}, \citenamefont {Huggins},
  \citenamefont {Ioffe}, \citenamefont {Isakov}, \citenamefont {Jeffrey},
  \citenamefont {Jiang}, \citenamefont {Jones}, \citenamefont {Kafri},
  \citenamefont {Kechedzhi}, \citenamefont {Kelly}, \citenamefont {Kim},
  \citenamefont {Klimov}, \citenamefont {Korotkov}, \citenamefont {Kostritsa},
  \citenamefont {Landhuis}, \citenamefont {Laptev}, \citenamefont {Lindmark},
  \citenamefont {Lucero}, \citenamefont {Martin}, \citenamefont {Martinis},
  \citenamefont {McClean}, \citenamefont {McEwen}, \citenamefont {Megrant},
  \citenamefont {Mi}, \citenamefont {Mohseni}, \citenamefont {Mruczkiewicz},
  \citenamefont {Mutus}, \citenamefont {Naaman}, \citenamefont {Neeley},
  \citenamefont {Neill}, \citenamefont {Neven}, \citenamefont {Niu},
  \citenamefont {O’Brien}, \citenamefont {Ostby}, \citenamefont {Petukhov},
  \citenamefont {Putterman}, \citenamefont {Quintana}, \citenamefont {Roushan},
  \citenamefont {Rubin}, \citenamefont {Sank}, \citenamefont {Satzinger},
  \citenamefont {Smelyanskiy}, \citenamefont {Strain}, \citenamefont {Sung},
  \citenamefont {Szalay}, \citenamefont {Takeshita}, \citenamefont
  {Vainsencher}, \citenamefont {White}, \citenamefont {Wiebe}, \citenamefont
  {Yao}, \citenamefont {Yeh},\ and\ \citenamefont
  {Zalcman}}]{ET_2020_SCIENCE_Zalcman}%
  \BibitemOpen
  \bibfield  {author} {\bibinfo {author} {\bibfnamefont {G.~A.}\ \bibnamefont
  {QUANTUM}}, \bibinfo {author} {\bibnamefont {COLLABORATORS}}, \bibinfo
  {author} {\bibfnamefont {F.}~\bibnamefont {Arute}}, \bibinfo {author}
  {\bibfnamefont {K.}~\bibnamefont {Arya}}, \bibinfo {author} {\bibfnamefont
  {R.}~\bibnamefont {Babbush}}, \bibinfo {author} {\bibfnamefont
  {D.}~\bibnamefont {Bacon}}, \bibinfo {author} {\bibfnamefont {J.~C.}\
  \bibnamefont {Bardin}}, \bibinfo {author} {\bibfnamefont {R.}~\bibnamefont
  {Barends}}, \bibinfo {author} {\bibfnamefont {S.}~\bibnamefont {Boixo}},
  \bibinfo {author} {\bibfnamefont {M.}~\bibnamefont {Broughton}}, \bibinfo
  {author} {\bibfnamefont {B.~B.}\ \bibnamefont {Buckley}}, \bibinfo {author}
  {\bibfnamefont {D.~A.}\ \bibnamefont {Buell}}, \bibinfo {author}
  {\bibfnamefont {B.}~\bibnamefont {Burkett}}, \bibinfo {author} {\bibfnamefont
  {N.}~\bibnamefont {Bushnell}}, \bibinfo {author} {\bibfnamefont
  {Y.}~\bibnamefont {Chen}}, \bibinfo {author} {\bibfnamefont {Z.}~\bibnamefont
  {Chen}}, \bibinfo {author} {\bibfnamefont {B.}~\bibnamefont {Chiaro}},
  \bibinfo {author} {\bibfnamefont {R.}~\bibnamefont {Collins}}, \bibinfo
  {author} {\bibfnamefont {W.}~\bibnamefont {Courtney}}, \bibinfo {author}
  {\bibfnamefont {S.}~\bibnamefont {Demura}}, \bibinfo {author} {\bibfnamefont
  {A.}~\bibnamefont {Dunsworth}}, \bibinfo {author} {\bibfnamefont
  {E.}~\bibnamefont {Farhi}}, \bibinfo {author} {\bibfnamefont
  {A.}~\bibnamefont {Fowler}}, \bibinfo {author} {\bibfnamefont
  {B.}~\bibnamefont {Foxen}}, \bibinfo {author} {\bibfnamefont
  {C.}~\bibnamefont {Gidney}}, \bibinfo {author} {\bibfnamefont
  {M.}~\bibnamefont {Giustina}}, \bibinfo {author} {\bibfnamefont
  {R.}~\bibnamefont {Graff}}, \bibinfo {author} {\bibfnamefont
  {S.}~\bibnamefont {Habegger}}, \bibinfo {author} {\bibfnamefont {M.~P.}\
  \bibnamefont {Harrigan}}, \bibinfo {author} {\bibfnamefont {A.}~\bibnamefont
  {Ho}}, \bibinfo {author} {\bibfnamefont {S.}~\bibnamefont {Hong}}, \bibinfo
  {author} {\bibfnamefont {T.}~\bibnamefont {Huang}}, \bibinfo {author}
  {\bibfnamefont {W.~J.}\ \bibnamefont {Huggins}}, \bibinfo {author}
  {\bibfnamefont {L.}~\bibnamefont {Ioffe}}, \bibinfo {author} {\bibfnamefont
  {S.~V.}\ \bibnamefont {Isakov}}, \bibinfo {author} {\bibfnamefont
  {E.}~\bibnamefont {Jeffrey}}, \bibinfo {author} {\bibfnamefont
  {Z.}~\bibnamefont {Jiang}}, \bibinfo {author} {\bibfnamefont
  {C.}~\bibnamefont {Jones}}, \bibinfo {author} {\bibfnamefont
  {D.}~\bibnamefont {Kafri}}, \bibinfo {author} {\bibfnamefont
  {K.}~\bibnamefont {Kechedzhi}}, \bibinfo {author} {\bibfnamefont
  {J.}~\bibnamefont {Kelly}}, \bibinfo {author} {\bibfnamefont
  {S.}~\bibnamefont {Kim}}, \bibinfo {author} {\bibfnamefont {P.~V.}\
  \bibnamefont {Klimov}}, \bibinfo {author} {\bibfnamefont {A.}~\bibnamefont
  {Korotkov}}, \bibinfo {author} {\bibfnamefont {F.}~\bibnamefont {Kostritsa}},
  \bibinfo {author} {\bibfnamefont {D.}~\bibnamefont {Landhuis}}, \bibinfo
  {author} {\bibfnamefont {P.}~\bibnamefont {Laptev}}, \bibinfo {author}
  {\bibfnamefont {M.}~\bibnamefont {Lindmark}}, \bibinfo {author}
  {\bibfnamefont {E.}~\bibnamefont {Lucero}}, \bibinfo {author} {\bibfnamefont
  {O.}~\bibnamefont {Martin}}, \bibinfo {author} {\bibfnamefont {J.~M.}\
  \bibnamefont {Martinis}}, \bibinfo {author} {\bibfnamefont {J.~R.}\
  \bibnamefont {McClean}}, \bibinfo {author} {\bibfnamefont {M.}~\bibnamefont
  {McEwen}}, \bibinfo {author} {\bibfnamefont {A.}~\bibnamefont {Megrant}},
  \bibinfo {author} {\bibfnamefont {X.}~\bibnamefont {Mi}}, \bibinfo {author}
  {\bibfnamefont {M.}~\bibnamefont {Mohseni}}, \bibinfo {author} {\bibfnamefont
  {W.}~\bibnamefont {Mruczkiewicz}}, \bibinfo {author} {\bibfnamefont
  {J.}~\bibnamefont {Mutus}}, \bibinfo {author} {\bibfnamefont
  {O.}~\bibnamefont {Naaman}}, \bibinfo {author} {\bibfnamefont
  {M.}~\bibnamefont {Neeley}}, \bibinfo {author} {\bibfnamefont
  {C.}~\bibnamefont {Neill}}, \bibinfo {author} {\bibfnamefont
  {H.}~\bibnamefont {Neven}}, \bibinfo {author} {\bibfnamefont {M.~Y.}\
  \bibnamefont {Niu}}, \bibinfo {author} {\bibfnamefont {T.~E.}\ \bibnamefont
  {O’Brien}}, \bibinfo {author} {\bibfnamefont {E.}~\bibnamefont {Ostby}},
  \bibinfo {author} {\bibfnamefont {A.}~\bibnamefont {Petukhov}}, \bibinfo
  {author} {\bibfnamefont {H.}~\bibnamefont {Putterman}}, \bibinfo {author}
  {\bibfnamefont {C.}~\bibnamefont {Quintana}}, \bibinfo {author}
  {\bibfnamefont {P.}~\bibnamefont {Roushan}}, \bibinfo {author} {\bibfnamefont
  {N.~C.}\ \bibnamefont {Rubin}}, \bibinfo {author} {\bibfnamefont
  {D.}~\bibnamefont {Sank}}, \bibinfo {author} {\bibfnamefont {K.~J.}\
  \bibnamefont {Satzinger}}, \bibinfo {author} {\bibfnamefont {V.}~\bibnamefont
  {Smelyanskiy}}, \bibinfo {author} {\bibfnamefont {D.}~\bibnamefont {Strain}},
  \bibinfo {author} {\bibfnamefont {K.~J.}\ \bibnamefont {Sung}}, \bibinfo
  {author} {\bibfnamefont {M.}~\bibnamefont {Szalay}}, \bibinfo {author}
  {\bibfnamefont {T.~Y.}\ \bibnamefont {Takeshita}}, \bibinfo {author}
  {\bibfnamefont {A.}~\bibnamefont {Vainsencher}}, \bibinfo {author}
  {\bibfnamefont {T.}~\bibnamefont {White}}, \bibinfo {author} {\bibfnamefont
  {N.}~\bibnamefont {Wiebe}}, \bibinfo {author} {\bibfnamefont {Z.~J.}\
  \bibnamefont {Yao}}, \bibinfo {author} {\bibfnamefont {P.}~\bibnamefont
  {Yeh}}, \ and\ \bibinfo {author} {\bibfnamefont {A.}~\bibnamefont
  {Zalcman}},\ }\href {\doibase 10.1126/science.abb9811} {\bibfield  {journal}
  {\bibinfo  {journal} {Science}\ }\textbf {\bibinfo {volume} {369}},\ \bibinfo
  {pages} {1084} (\bibinfo {year} {2020})}\BibitemShut {NoStop}%
\bibitem [{\citenamefont {Yan}\ \emph {et~al.}(2019)\citenamefont {Yan},
  \citenamefont {Zhang}, \citenamefont {Gong}, \citenamefont {Wu},
  \citenamefont {Zheng}, \citenamefont {Li}, \citenamefont {Wang},
  \citenamefont {Liang}, \citenamefont {Lin}, \citenamefont {Xu}, \citenamefont
  {Guo}, \citenamefont {Sun}, \citenamefont {Peng}, \citenamefont {Xia},
  \citenamefont {Deng}, \citenamefont {Rong}, \citenamefont {You},
  \citenamefont {Nori}, \citenamefont {Fan}, \citenamefont {Zhu},\ and\
  \citenamefont {Pan}}]{ET_2019_SCIENCE_Pan}%
  \BibitemOpen
  \bibfield  {author} {\bibinfo {author} {\bibfnamefont {Z.}~\bibnamefont
  {Yan}}, \bibinfo {author} {\bibfnamefont {Y.-R.}\ \bibnamefont {Zhang}},
  \bibinfo {author} {\bibfnamefont {M.}~\bibnamefont {Gong}}, \bibinfo {author}
  {\bibfnamefont {Y.}~\bibnamefont {Wu}}, \bibinfo {author} {\bibfnamefont
  {Y.}~\bibnamefont {Zheng}}, \bibinfo {author} {\bibfnamefont
  {S.}~\bibnamefont {Li}}, \bibinfo {author} {\bibfnamefont {C.}~\bibnamefont
  {Wang}}, \bibinfo {author} {\bibfnamefont {F.}~\bibnamefont {Liang}},
  \bibinfo {author} {\bibfnamefont {J.}~\bibnamefont {Lin}}, \bibinfo {author}
  {\bibfnamefont {Y.}~\bibnamefont {Xu}}, \bibinfo {author} {\bibfnamefont
  {C.}~\bibnamefont {Guo}}, \bibinfo {author} {\bibfnamefont {L.}~\bibnamefont
  {Sun}}, \bibinfo {author} {\bibfnamefont {C.-Z.}\ \bibnamefont {Peng}},
  \bibinfo {author} {\bibfnamefont {K.}~\bibnamefont {Xia}}, \bibinfo {author}
  {\bibfnamefont {H.}~\bibnamefont {Deng}}, \bibinfo {author} {\bibfnamefont
  {H.}~\bibnamefont {Rong}}, \bibinfo {author} {\bibfnamefont {J.~Q.}\
  \bibnamefont {You}}, \bibinfo {author} {\bibfnamefont {F.}~\bibnamefont
  {Nori}}, \bibinfo {author} {\bibfnamefont {H.}~\bibnamefont {Fan}}, \bibinfo
  {author} {\bibfnamefont {X.}~\bibnamefont {Zhu}}, \ and\ \bibinfo {author}
  {\bibfnamefont {J.-W.}\ \bibnamefont {Pan}},\ }\href {\doibase
  10.1126/science.aaw1611} {\bibfield  {journal} {\bibinfo  {journal}
  {Science}\ }\textbf {\bibinfo {volume} {364}},\ \bibinfo {pages} {753}
  (\bibinfo {year} {2019})}\BibitemShut {NoStop}%
\bibitem [{\citenamefont {Alicki}\ and\ \citenamefont
  {Fannes}(2013)}]{T_2013_PRE_Fannes}%
  \BibitemOpen
  \bibfield  {author} {\bibinfo {author} {\bibfnamefont {R.}~\bibnamefont
  {Alicki}}\ and\ \bibinfo {author} {\bibfnamefont {M.}~\bibnamefont
  {Fannes}},\ }\href {\doibase 10.1103/PhysRevE.87.042123} {\bibfield
  {journal} {\bibinfo  {journal} {Phys. Rev. E}\ }\textbf {\bibinfo {volume}
  {87}},\ \bibinfo {pages} {042123} (\bibinfo {year} {2013})}\BibitemShut
  {NoStop}%
\bibitem [{\citenamefont {Binder}\ \emph {et~al.}(2015)\citenamefont {Binder},
  \citenamefont {Vinjanampathy}, \citenamefont {Modi},\ and\ \citenamefont
  {Goold}}]{E_2015_NJP_Binder}%
  \BibitemOpen
  \bibfield  {author} {\bibinfo {author} {\bibfnamefont {F.~C.}\ \bibnamefont
  {Binder}}, \bibinfo {author} {\bibfnamefont {S.}~\bibnamefont
  {Vinjanampathy}}, \bibinfo {author} {\bibfnamefont {K.}~\bibnamefont {Modi}},
  \ and\ \bibinfo {author} {\bibfnamefont {J.}~\bibnamefont {Goold}},\ }\href
  {\doibase 10.1088/1367-2630/17/7/075015} {\bibfield  {journal} {\bibinfo
  {journal} {New Journal of Physics}\ }\textbf {\bibinfo {volume} {17}},\
  \bibinfo {pages} {075015} (\bibinfo {year} {2015})}\BibitemShut {NoStop}%
\bibitem [{\citenamefont {Campaioli}\ \emph {et~al.}(2017)\citenamefont
  {Campaioli}, \citenamefont {Pollock}, \citenamefont {Binder}, \citenamefont
  {C\'eleri}, \citenamefont {Goold}, \citenamefont {Vinjanampathy},\ and\
  \citenamefont {Modi}}]{E_2017_PRL_Kavan}%
  \BibitemOpen
  \bibfield  {author} {\bibinfo {author} {\bibfnamefont {F.}~\bibnamefont
  {Campaioli}}, \bibinfo {author} {\bibfnamefont {F.~A.}\ \bibnamefont
  {Pollock}}, \bibinfo {author} {\bibfnamefont {F.~C.}\ \bibnamefont {Binder}},
  \bibinfo {author} {\bibfnamefont {L.}~\bibnamefont {C\'eleri}}, \bibinfo
  {author} {\bibfnamefont {J.}~\bibnamefont {Goold}}, \bibinfo {author}
  {\bibfnamefont {S.}~\bibnamefont {Vinjanampathy}}, \ and\ \bibinfo {author}
  {\bibfnamefont {K.}~\bibnamefont {Modi}},\ }\href {\doibase
  10.1103/PhysRevLett.118.150601} {\bibfield  {journal} {\bibinfo  {journal}
  {Phys. Rev. Lett.}\ }\textbf {\bibinfo {volume} {118}},\ \bibinfo {pages}
  {150601} (\bibinfo {year} {2017})}\BibitemShut {NoStop}%
\bibitem [{\citenamefont {Gyhm}\ \emph {et~al.}(2022)\citenamefont {Gyhm},
  \citenamefont {\ifmmode~\check{S}\else \v{S}\fi{}afr\'anek},\ and\
  \citenamefont {Rosa}}]{Gyhm2022}%
  \BibitemOpen
  \bibfield  {author} {\bibinfo {author} {\bibfnamefont {J.-Y.}\ \bibnamefont
  {Gyhm}}, \bibinfo {author} {\bibfnamefont {D.}~\bibnamefont
  {\ifmmode~\check{S}\else \v{S}\fi{}afr\'anek}}, \ and\ \bibinfo {author}
  {\bibfnamefont {D.}~\bibnamefont {Rosa}},\ }\href {\doibase
  10.1103/PhysRevLett.128.140501} {\bibfield  {journal} {\bibinfo  {journal}
  {Phys. Rev. Lett.}\ }\textbf {\bibinfo {volume} {128}},\ \bibinfo {pages}
  {140501} (\bibinfo {year} {2022})}\BibitemShut {NoStop}%
\bibitem [{\citenamefont {Le}\ \emph {et~al.}(2018)\citenamefont {Le},
  \citenamefont {Levinsen}, \citenamefont {Modi}, \citenamefont {Parish},\ and\
  \citenamefont {Pollock}}]{T_2017_PRA_Felix}%
  \BibitemOpen
  \bibfield  {author} {\bibinfo {author} {\bibfnamefont {T.~P.}\ \bibnamefont
  {Le}}, \bibinfo {author} {\bibfnamefont {J.}~\bibnamefont {Levinsen}},
  \bibinfo {author} {\bibfnamefont {K.}~\bibnamefont {Modi}}, \bibinfo {author}
  {\bibfnamefont {M.~M.}\ \bibnamefont {Parish}}, \ and\ \bibinfo {author}
  {\bibfnamefont {F.~A.}\ \bibnamefont {Pollock}},\ }\href {\doibase
  10.1103/PhysRevA.97.022106} {\bibfield  {journal} {\bibinfo  {journal} {Phys.
  Rev. A}\ }\textbf {\bibinfo {volume} {97}},\ \bibinfo {pages} {022106}
  (\bibinfo {year} {2018})}\BibitemShut {NoStop}%
\bibitem [{\citenamefont {Ferraro}\ \emph {et~al.}(2018)\citenamefont
  {Ferraro}, \citenamefont {Campisi}, \citenamefont {Andolina}, \citenamefont
  {Pellegrini},\ and\ \citenamefont {Polini}}]{T_2018_PRL_Marco}%
  \BibitemOpen
  \bibfield  {author} {\bibinfo {author} {\bibfnamefont {D.}~\bibnamefont
  {Ferraro}}, \bibinfo {author} {\bibfnamefont {M.}~\bibnamefont {Campisi}},
  \bibinfo {author} {\bibfnamefont {G.~M.}\ \bibnamefont {Andolina}}, \bibinfo
  {author} {\bibfnamefont {V.}~\bibnamefont {Pellegrini}}, \ and\ \bibinfo
  {author} {\bibfnamefont {M.}~\bibnamefont {Polini}},\ }\href {\doibase
  10.1103/PhysRevLett.120.117702} {\bibfield  {journal} {\bibinfo  {journal}
  {Phys. Rev. Lett.}\ }\textbf {\bibinfo {volume} {120}},\ \bibinfo {pages}
  {117702} (\bibinfo {year} {2018})}\BibitemShut {NoStop}%
\bibitem [{\citenamefont {Zhang}\ and\ \citenamefont
  {blaauboer}(2018)}]{T_2018_AXRIV_blaauboer}%
  \BibitemOpen
  \bibfield  {author} {\bibinfo {author} {\bibfnamefont {X.}~\bibnamefont
  {Zhang}}\ and\ \bibinfo {author} {\bibfnamefont {M.}~\bibnamefont
  {blaauboer}},\ }\href@noop {} {\enquote {\bibinfo {title} {Enhanced energy
  transfer in a dicke quantum battery},}\ } (\bibinfo {year} {2018}),\ \Eprint
  {http://arxiv.org/abs/1812.10139} {arXiv:1812.10139 [quant-ph]} \BibitemShut
  {NoStop}%
\bibitem [{\citenamefont {Andolina}\ \emph
  {et~al.}(2019{\natexlab{a}})\citenamefont {Andolina}, \citenamefont {Keck},
  \citenamefont {Mari}, \citenamefont {Giovannetti},\ and\ \citenamefont
  {Polini}}]{T_2019_PRB_Andolina2}%
  \BibitemOpen
  \bibfield  {author} {\bibinfo {author} {\bibfnamefont {G.~M.}\ \bibnamefont
  {Andolina}}, \bibinfo {author} {\bibfnamefont {M.}~\bibnamefont {Keck}},
  \bibinfo {author} {\bibfnamefont {A.}~\bibnamefont {Mari}}, \bibinfo {author}
  {\bibfnamefont {V.}~\bibnamefont {Giovannetti}}, \ and\ \bibinfo {author}
  {\bibfnamefont {M.}~\bibnamefont {Polini}},\ }\href {\doibase
  10.1103/PhysRevB.99.205437} {\bibfield  {journal} {\bibinfo  {journal} {Phys.
  Rev. B}\ }\textbf {\bibinfo {volume} {99}},\ \bibinfo {pages} {205437}
  (\bibinfo {year} {2019}{\natexlab{a}})}\BibitemShut {NoStop}%
\bibitem [{\citenamefont {Andolina}\ \emph
  {et~al.}(2019{\natexlab{b}})\citenamefont {Andolina}, \citenamefont {Keck},
  \citenamefont {Mari}, \citenamefont {Campisi}, \citenamefont {Giovannetti},\
  and\ \citenamefont {Polini}}]{T_2019_PRL_Andolina3}%
  \BibitemOpen
  \bibfield  {author} {\bibinfo {author} {\bibfnamefont {G.~M.}\ \bibnamefont
  {Andolina}}, \bibinfo {author} {\bibfnamefont {M.}~\bibnamefont {Keck}},
  \bibinfo {author} {\bibfnamefont {A.}~\bibnamefont {Mari}}, \bibinfo {author}
  {\bibfnamefont {M.}~\bibnamefont {Campisi}}, \bibinfo {author} {\bibfnamefont
  {V.}~\bibnamefont {Giovannetti}}, \ and\ \bibinfo {author} {\bibfnamefont
  {M.}~\bibnamefont {Polini}},\ }\href {\doibase
  10.1103/PhysRevLett.122.047702} {\bibfield  {journal} {\bibinfo  {journal}
  {Phys. Rev. Lett.}\ }\textbf {\bibinfo {volume} {122}},\ \bibinfo {pages}
  {047702} (\bibinfo {year} {2019}{\natexlab{b}})}\BibitemShut {NoStop}%
\bibitem [{\citenamefont {Rossini}\ \emph {et~al.}(2019)\citenamefont
  {Rossini}, \citenamefont {Andolina},\ and\ \citenamefont
  {Polini}}]{T_2019_PRB_Polini}%
  \BibitemOpen
  \bibfield  {author} {\bibinfo {author} {\bibfnamefont {D.}~\bibnamefont
  {Rossini}}, \bibinfo {author} {\bibfnamefont {G.~M.}\ \bibnamefont
  {Andolina}}, \ and\ \bibinfo {author} {\bibfnamefont {M.}~\bibnamefont
  {Polini}},\ }\href {\doibase 10.1103/PhysRevB.100.115142} {\bibfield
  {journal} {\bibinfo  {journal} {Phys. Rev. B}\ }\textbf {\bibinfo {volume}
  {100}},\ \bibinfo {pages} {115142} (\bibinfo {year} {2019})}\BibitemShut
  {NoStop}%
\bibitem [{\citenamefont {Juli\`a-Farr\'e}\ \emph {et~al.}(2020)\citenamefont
  {Juli\`a-Farr\'e}, \citenamefont {Salamon}, \citenamefont {Riera},
  \citenamefont {Bera},\ and\ \citenamefont {Lewenstein}}]{T_2020_PRR_Maciej}%
  \BibitemOpen
  \bibfield  {author} {\bibinfo {author} {\bibfnamefont {S.}~\bibnamefont
  {Juli\`a-Farr\'e}}, \bibinfo {author} {\bibfnamefont {T.}~\bibnamefont
  {Salamon}}, \bibinfo {author} {\bibfnamefont {A.}~\bibnamefont {Riera}},
  \bibinfo {author} {\bibfnamefont {M.~N.}\ \bibnamefont {Bera}}, \ and\
  \bibinfo {author} {\bibfnamefont {M.}~\bibnamefont {Lewenstein}},\ }\href
  {\doibase 10.1103/PhysRevResearch.2.023113} {\bibfield  {journal} {\bibinfo
  {journal} {Phys. Rev. Research}\ }\textbf {\bibinfo {volume} {2}},\ \bibinfo
  {pages} {023113} (\bibinfo {year} {2020})}\BibitemShut {NoStop}%
\bibitem [{\citenamefont {Rossini}\ \emph {et~al.}(2020)\citenamefont
  {Rossini}, \citenamefont {Andolina}, \citenamefont {Rosa}, \citenamefont
  {Carrega},\ and\ \citenamefont {Polini}}]{T_2020_PRL_Polini}%
  \BibitemOpen
  \bibfield  {author} {\bibinfo {author} {\bibfnamefont {D.}~\bibnamefont
  {Rossini}}, \bibinfo {author} {\bibfnamefont {G.~M.}\ \bibnamefont
  {Andolina}}, \bibinfo {author} {\bibfnamefont {D.}~\bibnamefont {Rosa}},
  \bibinfo {author} {\bibfnamefont {M.}~\bibnamefont {Carrega}}, \ and\
  \bibinfo {author} {\bibfnamefont {M.}~\bibnamefont {Polini}},\ }\href
  {\doibase 10.1103/PhysRevLett.125.236402} {\bibfield  {journal} {\bibinfo
  {journal} {Phys. Rev. Lett.}\ }\textbf {\bibinfo {volume} {125}},\ \bibinfo
  {pages} {236402} (\bibinfo {year} {2020})}\BibitemShut {NoStop}%
\bibitem [{\citenamefont {Farina}\ \emph {et~al.}(2019)\citenamefont {Farina},
  \citenamefont {Andolina}, \citenamefont {Mari}, \citenamefont {Polini},\ and\
  \citenamefont {Giovannetti}}]{T_2019_PRB_Farina}%
  \BibitemOpen
  \bibfield  {author} {\bibinfo {author} {\bibfnamefont {D.}~\bibnamefont
  {Farina}}, \bibinfo {author} {\bibfnamefont {G.~M.}\ \bibnamefont
  {Andolina}}, \bibinfo {author} {\bibfnamefont {A.}~\bibnamefont {Mari}},
  \bibinfo {author} {\bibfnamefont {M.}~\bibnamefont {Polini}}, \ and\ \bibinfo
  {author} {\bibfnamefont {V.}~\bibnamefont {Giovannetti}},\ }\href {\doibase
  10.1103/PhysRevB.99.035421} {\bibfield  {journal} {\bibinfo  {journal} {Phys.
  Rev. B}\ }\textbf {\bibinfo {volume} {99}},\ \bibinfo {pages} {035421}
  (\bibinfo {year} {2019})}\BibitemShut {NoStop}%
\bibitem [{\citenamefont {Barra}(2019)}]{T_2019_PRL_Felipe}%
  \BibitemOpen
  \bibfield  {author} {\bibinfo {author} {\bibfnamefont {F.}~\bibnamefont
  {Barra}},\ }\href {\doibase 10.1103/PhysRevLett.122.210601} {\bibfield
  {journal} {\bibinfo  {journal} {Phys. Rev. Lett.}\ }\textbf {\bibinfo
  {volume} {122}},\ \bibinfo {pages} {210601} (\bibinfo {year}
  {2019})}\BibitemShut {NoStop}%
\bibitem [{\citenamefont {Crescente}\ \emph {et~al.}(2020)\citenamefont
  {Crescente}, \citenamefont {Carrega}, \citenamefont {Sassetti},\ and\
  \citenamefont {Ferraro}}]{T_2020_NJP_Ferraro}%
  \BibitemOpen
  \bibfield  {author} {\bibinfo {author} {\bibfnamefont {A.}~\bibnamefont
  {Crescente}}, \bibinfo {author} {\bibfnamefont {M.}~\bibnamefont {Carrega}},
  \bibinfo {author} {\bibfnamefont {M.}~\bibnamefont {Sassetti}}, \ and\
  \bibinfo {author} {\bibfnamefont {D.}~\bibnamefont {Ferraro}},\ }\href
  {\doibase 10.1088/1367-2630/ab91fc} {\bibfield  {journal} {\bibinfo
  {journal} {New Journal of Physics}\ }\textbf {\bibinfo {volume} {22}},\
  \bibinfo {pages} {063057} (\bibinfo {year} {2020})}\BibitemShut {NoStop}%
\bibitem [{\citenamefont {Carrega}\ \emph {et~al.}(2020)\citenamefont
  {Carrega}, \citenamefont {Crescente}, \citenamefont {Ferraro},\ and\
  \citenamefont {Sassetti}}]{T_2020_NJP_Sassetti}%
  \BibitemOpen
  \bibfield  {author} {\bibinfo {author} {\bibfnamefont {M.}~\bibnamefont
  {Carrega}}, \bibinfo {author} {\bibfnamefont {A.}~\bibnamefont {Crescente}},
  \bibinfo {author} {\bibfnamefont {D.}~\bibnamefont {Ferraro}}, \ and\
  \bibinfo {author} {\bibfnamefont {M.}~\bibnamefont {Sassetti}},\ }\href
  {\doibase 10.1088/1367-2630/abaa01} {\bibfield  {journal} {\bibinfo
  {journal} {New Journal of Physics}\ }\textbf {\bibinfo {volume} {22}},\
  \bibinfo {pages} {083085} (\bibinfo {year} {2020})}\BibitemShut {NoStop}%
\bibitem [{\citenamefont {Bai}\ and\ \citenamefont
  {An}(2020)}]{T_2020_PRA_Jun-Hong}%
  \BibitemOpen
  \bibfield  {author} {\bibinfo {author} {\bibfnamefont {S.-Y.}\ \bibnamefont
  {Bai}}\ and\ \bibinfo {author} {\bibfnamefont {J.-H.}\ \bibnamefont {An}},\
  }\href {\doibase 10.1103/PhysRevA.102.060201} {\bibfield  {journal} {\bibinfo
   {journal} {Phys. Rev. A}\ }\textbf {\bibinfo {volume} {102}},\ \bibinfo
  {pages} {060201} (\bibinfo {year} {2020})}\BibitemShut {NoStop}%
\bibitem [{\citenamefont {Garc\'{\i}a-Pintos}\ \emph
  {et~al.}(2020)\citenamefont {Garc\'{\i}a-Pintos}, \citenamefont {Hamma},\
  and\ \citenamefont {del Campo}}]{T_2020_PRL_Adolfo}%
  \BibitemOpen
  \bibfield  {author} {\bibinfo {author} {\bibfnamefont {L.~P.}\ \bibnamefont
  {Garc\'{\i}a-Pintos}}, \bibinfo {author} {\bibfnamefont {A.}~\bibnamefont
  {Hamma}}, \ and\ \bibinfo {author} {\bibfnamefont {A.}~\bibnamefont {del
  Campo}},\ }\href {\doibase 10.1103/PhysRevLett.125.040601} {\bibfield
  {journal} {\bibinfo  {journal} {Phys. Rev. Lett.}\ }\textbf {\bibinfo
  {volume} {125}},\ \bibinfo {pages} {040601} (\bibinfo {year}
  {2020})}\BibitemShut {NoStop}%
\bibitem [{\citenamefont {Quach}\ and\ \citenamefont
  {Munro}(2020)}]{T_2020_PRApplied_Quach}%
  \BibitemOpen
  \bibfield  {author} {\bibinfo {author} {\bibfnamefont {J.~Q.}\ \bibnamefont
  {Quach}}\ and\ \bibinfo {author} {\bibfnamefont {W.~J.}\ \bibnamefont
  {Munro}},\ }\href {\doibase 10.1103/PhysRevApplied.14.024092} {\bibfield
  {journal} {\bibinfo  {journal} {Phys. Rev. Applied}\ }\textbf {\bibinfo
  {volume} {14}},\ \bibinfo {pages} {024092} (\bibinfo {year}
  {2020})}\BibitemShut {NoStop}%
\bibitem [{\citenamefont {Ghosh}\ \emph {et~al.}(2021)\citenamefont {Ghosh},
  \citenamefont {Chanda}, \citenamefont {Mal},\ and\ \citenamefont
  {Sen(De)}}]{T_2021_PRA_Aditi}%
  \BibitemOpen
  \bibfield  {author} {\bibinfo {author} {\bibfnamefont {S.}~\bibnamefont
  {Ghosh}}, \bibinfo {author} {\bibfnamefont {T.}~\bibnamefont {Chanda}},
  \bibinfo {author} {\bibfnamefont {S.}~\bibnamefont {Mal}}, \ and\ \bibinfo
  {author} {\bibfnamefont {A.}~\bibnamefont {Sen(De)}},\ }\href {\doibase
  10.1103/PhysRevA.104.032207} {\bibfield  {journal} {\bibinfo  {journal}
  {Phys. Rev. A}\ }\textbf {\bibinfo {volume} {104}},\ \bibinfo {pages}
  {032207} (\bibinfo {year} {2021})}\BibitemShut {NoStop}%
\bibitem [{\citenamefont {Hu}\ \emph {et~al.}(2021)\citenamefont {Hu},
  \citenamefont {Qiu}, \citenamefont {Souza}, \citenamefont {Yuan},
  \citenamefont {Zhou}, \citenamefont {Zhang}, \citenamefont {Chu},
  \citenamefont {Pan}, \citenamefont {Hu}, \citenamefont {Li}, \citenamefont
  {Xu}, \citenamefont {Zhong}, \citenamefont {Liu}, \citenamefont {Yan},
  \citenamefont {Tan}, \citenamefont {Bachelard}, \citenamefont {Villas-Boas},
  \citenamefont {Santos},\ and\ \citenamefont {Yu}}]{E_2021_AXRIV_Yu}%
  \BibitemOpen
  \bibfield  {author} {\bibinfo {author} {\bibfnamefont {C.-K.}\ \bibnamefont
  {Hu}}, \bibinfo {author} {\bibfnamefont {J.}~\bibnamefont {Qiu}}, \bibinfo
  {author} {\bibfnamefont {P.~J.~P.}\ \bibnamefont {Souza}}, \bibinfo {author}
  {\bibfnamefont {J.}~\bibnamefont {Yuan}}, \bibinfo {author} {\bibfnamefont
  {Y.}~\bibnamefont {Zhou}}, \bibinfo {author} {\bibfnamefont {L.}~\bibnamefont
  {Zhang}}, \bibinfo {author} {\bibfnamefont {J.}~\bibnamefont {Chu}}, \bibinfo
  {author} {\bibfnamefont {X.}~\bibnamefont {Pan}}, \bibinfo {author}
  {\bibfnamefont {L.}~\bibnamefont {Hu}}, \bibinfo {author} {\bibfnamefont
  {J.}~\bibnamefont {Li}}, \bibinfo {author} {\bibfnamefont {Y.}~\bibnamefont
  {Xu}}, \bibinfo {author} {\bibfnamefont {Y.}~\bibnamefont {Zhong}}, \bibinfo
  {author} {\bibfnamefont {S.}~\bibnamefont {Liu}}, \bibinfo {author}
  {\bibfnamefont {F.}~\bibnamefont {Yan}}, \bibinfo {author} {\bibfnamefont
  {D.}~\bibnamefont {Tan}}, \bibinfo {author} {\bibfnamefont {R.}~\bibnamefont
  {Bachelard}}, \bibinfo {author} {\bibfnamefont {C.~J.}\ \bibnamefont
  {Villas-Boas}}, \bibinfo {author} {\bibfnamefont {A.~C.}\ \bibnamefont
  {Santos}}, \ and\ \bibinfo {author} {\bibfnamefont {D.}~\bibnamefont {Yu}},\
  }\href@noop {} {\enquote {\bibinfo {title} {Optimal charging of a
  superconducting quantum battery},}\ } (\bibinfo {year} {2021}),\ \Eprint
  {http://arxiv.org/abs/2108.04298} {arXiv:2108.04298 [quant-ph]} \BibitemShut
  {NoStop}%
\bibitem [{\citenamefont {Quach}\ \emph {et~al.}(2022)\citenamefont {Quach},
  \citenamefont {McGhee}, \citenamefont {Ganzer}, \citenamefont {Rouse},
  \citenamefont {Lovett}, \citenamefont {Gauger}, \citenamefont {Keeling},
  \citenamefont {Cerullo}, \citenamefont {Lidzey},\ and\ \citenamefont
  {Virgili}}]{Virgili2022}%
  \BibitemOpen
  \bibfield  {author} {\bibinfo {author} {\bibfnamefont {J.~Q.}\ \bibnamefont
  {Quach}}, \bibinfo {author} {\bibfnamefont {K.~E.}\ \bibnamefont {McGhee}},
  \bibinfo {author} {\bibfnamefont {L.}~\bibnamefont {Ganzer}}, \bibinfo
  {author} {\bibfnamefont {D.~M.}\ \bibnamefont {Rouse}}, \bibinfo {author}
  {\bibfnamefont {B.~W.}\ \bibnamefont {Lovett}}, \bibinfo {author}
  {\bibfnamefont {E.~M.}\ \bibnamefont {Gauger}}, \bibinfo {author}
  {\bibfnamefont {J.}~\bibnamefont {Keeling}}, \bibinfo {author} {\bibfnamefont
  {G.}~\bibnamefont {Cerullo}}, \bibinfo {author} {\bibfnamefont {D.~G.}\
  \bibnamefont {Lidzey}}, \ and\ \bibinfo {author} {\bibfnamefont
  {T.}~\bibnamefont {Virgili}},\ }\href {\doibase 10.1126/sciadv.abk3160}
  {\bibfield  {journal} {\bibinfo  {journal} {Science Advances}\ }\textbf
  {\bibinfo {volume} {8}},\ \bibinfo {pages} {eabk3160} (\bibinfo {year}
  {2022})},\ \Eprint
  {http://arxiv.org/abs/https://www.science.org/doi/pdf/10.1126/sciadv.abk3160}
  {https://www.science.org/doi/pdf/10.1126/sciadv.abk3160} \BibitemShut
  {NoStop}%
\bibitem [{\citenamefont {Andolina}\ \emph {et~al.}(2018)\citenamefont
  {Andolina}, \citenamefont {Farina}, \citenamefont {Mari}, \citenamefont
  {Pellegrini}, \citenamefont {Giovannetti},\ and\ \citenamefont
  {Polini}}]{T_2018_PRB_Andolina}%
  \BibitemOpen
  \bibfield  {author} {\bibinfo {author} {\bibfnamefont {G.~M.}\ \bibnamefont
  {Andolina}}, \bibinfo {author} {\bibfnamefont {D.}~\bibnamefont {Farina}},
  \bibinfo {author} {\bibfnamefont {A.}~\bibnamefont {Mari}}, \bibinfo {author}
  {\bibfnamefont {V.}~\bibnamefont {Pellegrini}}, \bibinfo {author}
  {\bibfnamefont {V.}~\bibnamefont {Giovannetti}}, \ and\ \bibinfo {author}
  {\bibfnamefont {M.}~\bibnamefont {Polini}},\ }\href {\doibase
  10.1103/PhysRevB.98.205423} {\bibfield  {journal} {\bibinfo  {journal} {Phys.
  Rev. B}\ }\textbf {\bibinfo {volume} {98}},\ \bibinfo {pages} {205423}
  (\bibinfo {year} {2018})}\BibitemShut {NoStop}%
\bibitem [{\citenamefont {Huangfu}\ and\ \citenamefont
  {Jing}(2021)}]{T_2021_PRE_Jing}%
  \BibitemOpen
  \bibfield  {author} {\bibinfo {author} {\bibfnamefont {Y.}~\bibnamefont
  {Huangfu}}\ and\ \bibinfo {author} {\bibfnamefont {J.}~\bibnamefont {Jing}},\
  }\href {\doibase 10.1103/PhysRevE.104.024129} {\bibfield  {journal} {\bibinfo
   {journal} {Phys. Rev. E}\ }\textbf {\bibinfo {volume} {104}},\ \bibinfo
  {pages} {024129} (\bibinfo {year} {2021})}\BibitemShut {NoStop}%
\bibitem [{\citenamefont {Xu}\ \emph {et~al.}(2020)\citenamefont {Xu},
  \citenamefont {Sun}, \citenamefont {Liu}, \citenamefont {Zhang},
  \citenamefont {Li}, \citenamefont {Dong}, \citenamefont {Ren}, \citenamefont
  {Zhang}, \citenamefont {Nori}, \citenamefont {Zheng}, \citenamefont {Fan},\
  and\ \citenamefont {Wang}}]{Xu2020}%
  \BibitemOpen
  \bibfield  {author} {\bibinfo {author} {\bibfnamefont {K.}~\bibnamefont
  {Xu}}, \bibinfo {author} {\bibfnamefont {Z.-H.}\ \bibnamefont {Sun}},
  \bibinfo {author} {\bibfnamefont {W.}~\bibnamefont {Liu}}, \bibinfo {author}
  {\bibfnamefont {Y.-R.}\ \bibnamefont {Zhang}}, \bibinfo {author}
  {\bibfnamefont {H.}~\bibnamefont {Li}}, \bibinfo {author} {\bibfnamefont
  {H.}~\bibnamefont {Dong}}, \bibinfo {author} {\bibfnamefont {W.}~\bibnamefont
  {Ren}}, \bibinfo {author} {\bibfnamefont {P.}~\bibnamefont {Zhang}}, \bibinfo
  {author} {\bibfnamefont {F.}~\bibnamefont {Nori}}, \bibinfo {author}
  {\bibfnamefont {D.}~\bibnamefont {Zheng}}, \bibinfo {author} {\bibfnamefont
  {H.}~\bibnamefont {Fan}}, \ and\ \bibinfo {author} {\bibfnamefont
  {H.}~\bibnamefont {Wang}},\ }\href {\doibase 10.1126/sciadv.aba4935}
  {\bibfield  {journal} {\bibinfo  {journal} {Science Advances}\ }\textbf
  {\bibinfo {volume} {6}},\ \bibinfo {pages} {eaba4935} (\bibinfo {year}
  {2020})}\BibitemShut {NoStop}%
\bibitem [{\citenamefont {Peng}\ \emph {et~al.}(2021)\citenamefont {Peng},
  \citenamefont {He}, \citenamefont {Chesi}, \citenamefont {Lin},\ and\
  \citenamefont {Guan}}]{T_2021_PRA_Guan}%
  \BibitemOpen
  \bibfield  {author} {\bibinfo {author} {\bibfnamefont {L.}~\bibnamefont
  {Peng}}, \bibinfo {author} {\bibfnamefont {W.-B.}\ \bibnamefont {He}},
  \bibinfo {author} {\bibfnamefont {S.}~\bibnamefont {Chesi}}, \bibinfo
  {author} {\bibfnamefont {H.-Q.}\ \bibnamefont {Lin}}, \ and\ \bibinfo
  {author} {\bibfnamefont {X.-W.}\ \bibnamefont {Guan}},\ }\href {\doibase
  10.1103/PhysRevA.103.052220} {\bibfield  {journal} {\bibinfo  {journal}
  {Phys. Rev. A}\ }\textbf {\bibinfo {volume} {103}},\ \bibinfo {pages}
  {052220} (\bibinfo {year} {2021})}\BibitemShut {NoStop}%
\bibitem [{\citenamefont {Bardarson}\ \emph {et~al.}(2012)\citenamefont
  {Bardarson}, \citenamefont {Pollmann},\ and\ \citenamefont
  {Moore}}]{Bardarson2012}%
  \BibitemOpen
  \bibfield  {author} {\bibinfo {author} {\bibfnamefont {J.~H.}\ \bibnamefont
  {Bardarson}}, \bibinfo {author} {\bibfnamefont {F.}~\bibnamefont {Pollmann}},
  \ and\ \bibinfo {author} {\bibfnamefont {J.~E.}\ \bibnamefont {Moore}},\
  }\href {\doibase 10.1103/PhysRevLett.109.017202} {\bibfield  {journal}
  {\bibinfo  {journal} {Phys. Rev. Lett.}\ }\textbf {\bibinfo {volume} {109}},\
  \bibinfo {pages} {017202} (\bibinfo {year} {2012})}\BibitemShut {NoStop}%
\bibitem [{\citenamefont {Serbyn}\ \emph {et~al.}(2013)\citenamefont {Serbyn},
  \citenamefont {Papi\ifmmode~\acute{c}\else \'{c}\fi{}},\ and\ \citenamefont
  {Abanin}}]{Serbyn2013}%
  \BibitemOpen
  \bibfield  {author} {\bibinfo {author} {\bibfnamefont {M.}~\bibnamefont
  {Serbyn}}, \bibinfo {author} {\bibfnamefont {Z.}~\bibnamefont
  {Papi\ifmmode~\acute{c}\else \'{c}\fi{}}}, \ and\ \bibinfo {author}
  {\bibfnamefont {D.~A.}\ \bibnamefont {Abanin}},\ }\href {\doibase
  10.1103/PhysRevLett.110.260601} {\bibfield  {journal} {\bibinfo  {journal}
  {Phys. Rev. Lett.}\ }\textbf {\bibinfo {volume} {110}},\ \bibinfo {pages}
  {260601} (\bibinfo {year} {2013})}\BibitemShut {NoStop}%
\end{thebibliography}%
\end{document}